\documentclass[aps,prd,superscriptaddress,amsmath,groupedaddress,reprint, nofootinbib, showpacs,floatfix, 10pt]{revtex4-2}
\usepackage{graphicx}
\usepackage{revsymb}
\usepackage{dcolumn}
\newcommand{\msb}{\mathrm{\overline{MS}}}
\newcommand{\almsb}{\overline{\alpha}_s}
\newcommand{\mmsb}{\overline{m}}
\newcommand{\eq}[1]{Eq.~(\ref{#1})}

\newcommand{\xv}{{\bf x}}
\newcommand{\pv}{{\bf p}}
\newcommand{\psib}{\overline{\psi}}

\newcommand{\e}{\mathrm{e}}

\newcommand{\nl}{\nonumber \\}
\newcommand{\Dv}{\mathbf{D}}
\newcommand{\Dvfour}{(\mathbf{D}^2)^2}
\newcommand{\Ev}{\mathbf{E}}
\newcommand{\Bv}{\mathbf{B}}
\newcommand{\Deltav}{\mathbf{\Delta}} 
\newcommand{\sigmav}{\mbox{\boldmath$\sigma$}}
\newcommand{\qed}{\mathrm{QED}}

\usepackage[colorlinks=true, linkcolor=blue, urlcolor=blue, citecolor=blue]{hyperref}
\usepackage{microtype}

\newcommand{\mbmbfour}{\rtag{4.1992(63)}}

\newcommand{\mbmbfive}{\rtag{4.1927(63)}}
\newcommand{\mbmbfourqed}{\rtag{4.1984(63)}}
\newcommand{\mbmbfiveqed}{\rtag{4.1923(63)}}
\newcommand{\mbtenfiveqed}{\rtag{3.6454(94)}}
\newcommand{\mbthreefourqed}{\rtag{4.500(10)}}
\newcommand{\mcthreefourqed}{\rtag{0.9813(34)}}
\newcommand{\mcmcfourqed}{\rtag{1.2716(82)}}
\newcommand{\msthreefourqed}{\rtag{83.39(26)}}
\newcommand{\msthreefourqedmean}{\rtag{83.39}}
\newcommand{\mstwofourqed}{\rtag{92.38(43)}}

\usepackage{times}

\usepackage{xcolor}
\newcommand{\rtag}[1]{#1}
\newcommand{\btag}[1]{#1}

\begin{document}
	\title{New high-precision $b$, $c$, and $s$ masses from pseudoscalar-pseudoscalar correlators\\
	     in $n_f=4$ lattice QCD}
    \author{Brian~\surname{Colquhoun}} 
    \affiliation{School of Physics and Astronomy, University of Glasgow, Glasgow, G12 8QQ, UK}

    \author{Christine~T.~H.~\surname{Davies}} 
    \email[]{Christine.Davies@glasgow.ac.uk}
    \affiliation{School of Physics and Astronomy, University of Glasgow, Glasgow, G12 8QQ, UK}

    \author{Daniel~\surname{Hatton}} 
    \affiliation{School of Physics and Astronomy, University of Glasgow, Glasgow, G12 8QQ, UK}

    \author{G.~Peter~\surname{Lepage}} 
    \email[]{g.p.lepage@cornell.edu}
	\affiliation{Laboratory for Elementary-Particle Physics,
		Cornell University, Ithaca, NY 14853, USA}

    \collaboration{HPQCD Collaboration}
    \homepage[URL: ]{https://www.physics.gla.ac.uk/hpqcd/}

    \date{April 1, 2026}

\begin{abstract}
We extend an earlier lattice QCD analysis of heavy-quark current-current correlators to obtain new values for the $\msb$ masses of the $b$, $c$, and $s$~quarks. The analysis uses gluon configurations from the MILC collaboration with vacuum polarization contributions from $u$, $d$, $s$, and~$c$ quarks ($n_f=4$), and lattice spacings down to~0.032~fm. {We find that $\mmsb_b(\mmsb_b, n_f=5)=\mbmbfiveqed$~GeV, $\mmsb_c(3~\mathrm{GeV}, n_f=4)=\mcthreefourqed$~GeV, and $\mmsb_s(3~\mathrm{GeV}, n_f=4)=\msthreefourqed$~MeV.} These results are corrected for QED by including (quenched) QED in the simulations. They are among the most accurate values by any method to date. We give a detailed analysis of finite lattice-spacing errors that shows why the HISQ discretization of the quark action is particularly useful for $b$-quark simulations even for lattices where~$am_b\approx1$. We also calculate QED and isospin corrections to the (fictitious) $\eta_s$-meson mass, which is used to tune $s$-quark masses in lattice simulations.
\end{abstract}

\maketitle
\section{Introduction}
Accurate values for heavy-quark masses are important for QCD phenomenology and indispensable for high-precision studies of the Higgs particle~\cite{Lepage:2014fla, Freitas:2019bre, deBlas:2019rxi}. 
The $b$-quark mass $m_b$ is particularly important for Higgs studies because $H\to b\overline b$ is the Higgs particle's dominant decay channel and its branching fraction is proportional to~$m_b^2$. In this paper we extract a new, very accurate value for the $b$~quark's mass from lattice simulations of heavy-quark current-current correlators. 

High-precision lattice simulations involving $b$~quarks are challenging because the large quark mass requires much smaller lattice spacings: for example, lattice errors proportional to $(am_h)^2$ are an order-of-magnitude larger for $b$ quarks (mass $m_h=m_b$) than for $c$~quarks ($m_h=m_c$). To achieve high precision, we employ the HISQ discretization of the quark Lagrangian~\cite{Follana:2006rc}, as this allows us to use much larger masses than other discretizations. In particular, as we show here, $am_h$~errors are suppressed by powers of $(v/c)^2$ when HISQ quarks are used to study nonrelativistic heavy-quark systems, where $v$~is the typical velocity of the heavy quark.

We determine the $b$~quark's lattice mass~$m_b$ by tuning it so that the lattice simulations reproduce the $\eta_b$~meson's mass obtained from experiment. The $\eta_b$ is nonrelativistice, and therefore $am_b$~errors are strongly suppressed when calculating its mass. We convert the lattice mass to the $\msb$ $b$-quark mass using lattice simulations to calculate moments
of the pseudoscalar correlator for a heavy quark with bare mass~$m_h$ (on the lattice)~\cite{HPQCD:2008kxl,McNeile:2010ji,Chakraborty:2014aca}:
\begin{align}
    G_n &\equiv \sum_t (t/a)^n G(t), \\
    G(t) &\equiv a^6 \sum_{\xv} (am_h)^2 \langle 0| j_5(\xv,t) j_5(0,0) | 0\rangle,
\end{align}
where $j_5\equiv \psib_h \gamma_5 \psi_h$ is the heavy quark's pseudoscalar density, $a$ is the lattice spacing, time $t$ is Euclidean and periodic with period T, and $\xv$ is the spatial position of the current. While $am_h$ errors are large for $m_h=m_b$ and small values of~$n$, they are suppressed as $n$ increases and the moments become nonrelativistic. 

We describe our lattice analysis in Sec.~\ref{sec:Lattice-Analysis}. Sec.~\ref{sec:moments} reviews how to extract the $\msb$ $b$-quark mass by combining lattice results for the correlator moments with results for the moments from continuum perturbation theory; it also shows how the moments become nonrelativistic as $n$~increases. Sec.~\ref{sec:finite-a} discusses finite-lattice-spacing errors and in particular $am_h$ errors for heavy quarks, showing how they are strongly suppressed for nonrelativistic systems when using the HISQ discretization of the quark action. This discussion builds on a detailed analysis of $am_h$ errors using tree-level perturbation theory in Appendices~\ref{app:nrHISQ} and~\ref{app:ghost}. Sec.~\ref{sec:a-uncertainty} discusses why $b$-quark masses are unusually insensitive to the accuracy with which the lattice spacing is known. We present details about the lattice simulations in Sec.~\ref{sec:simulations}, and describe how we analyze the lattice results in Secs.~\ref{sec:Rn-fit} and~\ref{sec:fit2}. 

We summarize our final results, before adding QED corrections, in Sec.~\ref{sec:qcd-results}, and review the (quenched) QED corrections, from earlier studies, in Sec.~\ref{sec:qed}. In Sec.~\ref{sec:charm} we use our new result for the $b$~mass, together with recent nonperturbative results for the ratio of $b$~and $c$~masses, to generate a new result for the $c$-quark mass. We discuss why this is a particularly accurate way to determine the $c$-quark mass. We use the same techique in Sec.~\ref{sec:smass} to obtain a new result for the $s$-quark mass. This relies upon a new lattice analysis of QED and isospin corrections to the mass of the fictitious $\eta_s$~meson, which is used to tune the $s$-quark mass in simulations (see Appendix~\ref{appendix:ms}). Finally we compare our determinations of the masses with other determinations, and summarize our conclusions in Sec.~\ref{sec:conclusions}.

\section{Lattice Analysis}
\label{sec:Lattice-Analysis}
The approach used  here to determine the $\msb$ $b$-quark mass $\mmsb_b(\mu)$ follows that used in~\cite{Chakraborty:2014aca} to determine the $c$-quark mass. We extend the previous analysis by adding results from simulations with a smaller lattice spacing~(0.032~fm), to accommodate the $b$~quark's larger mass. Our analysis of the $c$-quark mass used only moments with~$n=4$, 6, 8, and~10. Here we use all moments from $n=6$ through~$n=22$, which also helps suppress $am_b$ errors, as we show.

\subsection{Heavy-quark correlator moments}
\label{sec:moments}
As in~\cite{Chakraborty:2014aca}, we reduce finite-lattice-spacing, tuning and perturbation theory errors by introducing (dimensionless) reduced moments,\footnote{This formula had an extra factor of $1/m_c$ in \cite{Chakraborty:2014aca}. That factor would be $1/m_b$ in the current analysis. We do not include such a factor here because it leads to a D'Agostini bias in the final fit results~\cite{DAgostini:1993arp}. The bias is negligibly small but it can be avoided altogether by omitting the $1/m_b$ in this formula and adjusting the fit formula accordingly.}
\begin{equation}
    \label{eq:Rn}
    R_n(m_h) \equiv \Big(G_n / G^{(0)}_n\Big)^{1/(n-4)},
\end{equation}
where $G^{(0)}_n$ is the moment calculated on the same lattice but in lowest-order lattice perturbation theory.
We restrict our attention to moments with even $n\ge6$ as these are the only moments that depend strongly on the heavy~quark's mass~$m_h$.

The bare lattice quark mass $am_h$ for a heavy quark is related to the mass parameter $am_{0h}$ appearing in the HISQ Lagrangian by \eq{m0eqn} in Appendix~\ref{app:nrHISQ}. $am_h$ is the tree-level HISQ pole mass corrected to remove finite lattice-spacing errors to all orders in $(am_h)^2$. The difference between $am_0$ and $am$ is negligible for $u$, $d$, $s$, and $c$ quarks on the lattices we use; the difference is only~3\% even when~$am_h=1$. 

A lattice quark mass~$m_h$ is related to its $\msb$ mass $\mmsb_h(\mu)$ by a renormalization constant
\begin{equation}
    \label{eq:Zm}
    Z_m^\msb(\mu, a) \equiv \frac{\mmsb_h(\mu)}{m_h(a)}
\end{equation}
which is mass-independent up to finite lattice-spacing errors~\cite{Davies:2009ih,Hatton:2021syc}. This allows us to relate results calculated with various heavy-quark masses~$m_h$ to the $b$-quark mass since
\begin{equation}
    \label{eq:mmsb_m}
    \frac{\mmsb_h(\mu)}{m_h(a)} = \frac{\mmsb_b(\mu)}{m_b(a)},
\end{equation}
again up to finite lattice-spacing errors.

\begin{figure}
    \begin{center}
    \includegraphics[scale=0.9]{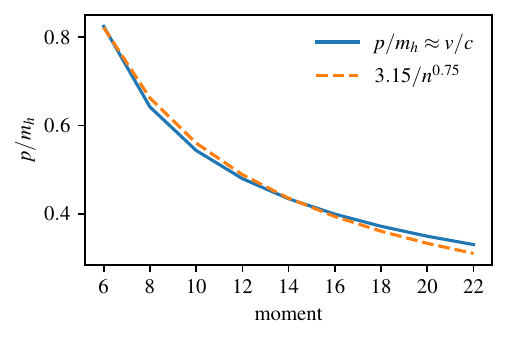}    
    \end{center}
\caption{\label{fig:pv_m} Average three momentum $p$, divided by the 
bare mass $m_h$, of the heavy valence quarks for
moment  $G_n$ in lowest-order perturbation theory (solid line). The average is  calculated 
by inserting a factor of $\log({\bf p}^2)/2$ into the Feynman integral
over Euclidean four-momentum~$p_\mu$\,\cite{Lepage:1992xa}. $p/m_h\approx v/c$, where $v$ is the typical heavy-quark velocity, is well approximated by function $3.15/n^{0.75}$ (dashed line).}
\end{figure}

The correlators are evaluated at zero energy, which is far below threshold: $E_\mathrm{th}=m_{\eta_h}$, where the ground-state pseudoscalar masses $m_{\eta_h}$ range from 6.6~to 9.7~GeV for the quark masses considered here. The heavy quarks also have large three-momenta, driving them even further off shell, when $n$~is small (see Fig.~\ref{fig:pv_m}). As a result we expect that the reduced moments are well described by perturbation theory, with 
\begin{align}
    \label{eq:Rnpth0}
    R_n(m_h) &= \frac{m_h}{\mmsb_h(\mu_n)}\, r_n\big(\almsb(\mu_n), \rho_n\big) \\
    &= \frac{m_b}{\mmsb_b(\mu_n)}\, r_n\big(\almsb(\mu_n), \rho_n\big)
    \label{eq:Rnpth}
\end{align}
up to finite lattice-spacing errors, where $\almsb(\mu_n)$ is the $\msb$ coupling at renormalization scale~$\mu_n$.\footnote{The mass factor in the perturbative expression for~$R_n(m_h)$ (\eq{eq:Rnpth0}) arises because, from its dimensions, moment $G_n \propto 1/\mathrm{mass}^{(n-4)}$ and the heavy valence quark mass is the dominant scale in the ultraviolet-finite moment. Specifically, the $m_h$ comes from $G_n^{(0)}$ in \eq{eq:Rn} evaluated in lattice perturbation theory while the $\mmsb_h(\mu_n)$ comes from $G_n$ evaluated in $\msb$ perturbation theory.}
Here constant~$\rho_n$ specifies the scale~$\mu_n$ in units of the mass:
\begin{equation}
    \mu_n \equiv \rho_n \mmsb_h(\mu_n).
\end{equation}
$r_n$~is a perturbation series, 
\begin{equation}
    \label{eq:rn-pth}
   r_n
   = 1 + \sum_{j=1}^\infty r_{nj}(\rho_n) \,\almsb^j(\mu_n),
\end{equation}
with the same anomalous dimension as $\mmsb_b(\mu_n)$, making $R_n(m_h)/m_b$ independent of both the lattice spacing (neglecting finite-$a$ errors) and the renormalization scale specified by~$\rho_n$. The dependence of the coefficients $r_{nj}$ on the sea-quark masses is implicit (and small). 

\eq{eq:Rnpth} can be rewritten
\begin{align}
    \label{eq:mbmu1}
\mmsb_b(\mu_n) &= m_b \, \frac{r_n(\almsb(\mu_n),\rho_n)}{R_n(m_h)},\\
\label{eq:mbmu2}
\mu_n &=  \rho_n\,\frac{am_h}{am_b}\,\mmsb_b(\mu_n) = \rho_n\mmsb_h(\mu_n).
\end{align}
Our strategy is to solve these equations (iteratively, for example) for $\mmsb_b(\mu_n)$ and $\mu_n$ given (nonperturbative) lattice values for $m_b$, $(am_h)/(am_b)$, and $R_n(m_h)$, and perturbation theory for~$r_n$.
From \eq{eq:Zm}, note that
\begin{equation}
Z_m^\msb(\mu_n, a) = \frac{r_n(\almsb(\mu_n),\rho_n)}{R_n(m_h(a))}
\end{equation}
up to finite lattice-spacing errors.

\begin{table}
    \caption{Perturbation theory coefficients for $r_n$ (\eq{eq:rn-pth}) with $n_f=4$ massless sea quarks. The coefficients are for renormalization scale $\mu_n$ wbere $\mu_n/\mmsb_h(\mu_n)$ is  specified in the second column and $\mmsb_h(\mu_n)$ is the heavy (valence) quark's $\msb$~mass. These coefficients are derived from results given in~\cite{Chetyrkin:2006xg,Boughezal:2006px,Maier:2008he,Kiyo:2009gb,Maier:2009fz}. Published results for the third coefficient when $n\ge 12$ include the valence quark in the sea, which is not done in the lattice simulations used here. We account for this difference by adding a 20\% uncertainty to the coefficients evaluated at $\mu/\mmsb_h=1$. This is based upon the behavior of the first four moments where the correction is of order 10\%~\rtag{\cite{Maier:2009fz}}.
    }
    \label{tab:rn-pth}
        \begin{ruledtabular}
            \begin{tabular}{rccccc}
            $n$ & $\rho_n\equiv\mu_n/\mmsb_h(\mu_n)$ & $r_{n1}$ & $r_{n2}$ & $r_{n3}$ \\
            \hline
            6 &  1.96 &  0.18758 &  0.46530 &   0.46324 \\ 
            8 &  1.48 &  0.06681 &  0.28741 &   0.28934 \\ 
           10 &  1.24 &  0.04917 &  0.23879 &   0.24074 \\ 
           12 &  1.17 &  0.00812 &  0.19308 & 0.192\,(62) \\ 
           14 &  1.08 &  0.00545 &  0.17207 & 0.171\,(48) \\ 
           16 &  1.02 &  0.00198 &  0.15385 & 0.156\,(35) \\ 
           18 &  0.97 &  0.00285 &  0.14025 & 0.142\,(25) \\ 
           20 &  0.93 &  0.00446 &  0.12883 & 0.129\,(17) \\ 
           22 &  0.90 &  0.00439 &  0.11765 & 0.120\,(10) \\ 
            \end{tabular}
        \end{ruledtabular}
    \end{table}

Values for the first three coefficients $r_{nj}$ in \eq{eq:rn-pth} are given in Table~\ref{tab:rn-pth}, for moments up to $n=22$~\cite{Chetyrkin:2006xg,Boughezal:2006px,Maier:2008he,Kiyo:2009gb,Maier:2009fz}. (Results through second order for larger values of $n$ can be found in~\cite{Maier:2007yn}.) These coefficients assume massless sea quarks, which is reasonable for $u$, $d$ and~$s$ quarks but less so for $c$~quarks. There are corrections of relative order $(m_c/m_h)^2$ to $r_{n2}$ and $r_{n3}$ but our simulations indicate that these are negligible. 

The perturbative coefficients are sensitive to the renormalization scale $\mu_n$ and become quite large if $\rho_n=\mu_n/\mmsb_h(\mu_n)$ is very large or small. We choose $\rho_n$ for the $n^\mathrm{th}$~moment to minimize the largest of the first three perturbative coefficients\,---\,that is, to minimize 
\begin{equation}
    \label{eq:rnmax}
    r_n^\mathrm{max} \equiv \mathrm{max}(|r_{n1}|, |r_{n2}|, |r_{n3}|).
\end{equation}
This criterion gives larger scales for smaller values of $n$ (Table~\ref{tab:rn-pth}), 
which is consistent with 
the larger quark three-momentum at small $n$ (Fig.~\ref{fig:pv_m}). 

The coefficients $r_{nj}$ are small, especially for larger~$n$. And the leading coefficients~$r_{n1}$ are very small. This suggests that perturbation theory will be quite convergent, especially if we restrict ourselves to large valence~quark masses, so that the perturbative scales, $\mu_n\propto\mmsb_h(\mu_n)$, are large and $\almsb(\mu_n)$ is small.

\subsection{Finite lattice-spacings errors}
\label{sec:finite-a}
\begin{figure}
    \begin{center}
        \includegraphics[scale=0.9]{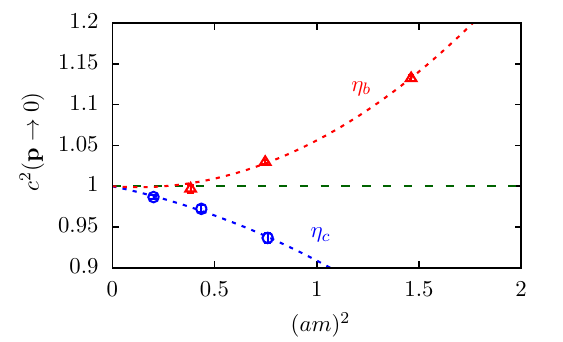}
    \end{center}
\caption{\label{fig:csq}Plots of $c^2(\pv\to 0)$ for the $\eta_b$ (red, triangles) and $\eta_c$ (blue, circles) mesons versus the valence-quark mass $am$ in lattice units. Fits of these data to a formula of the form $c_2 (am)^2 + c_4 (am)^4$ are also shown (short dashes). Results for the $\eta_b$ are from simulations with gluon ensembles~EF-5, UF-5, and~SF-5. Results for the $\eta_c$ are from coarser lattices, with lattice spacings of~0.09~fm, 0.12~fm and~0.15~fm. }
\end{figure}

There are three sources of finite lattice-spacing errors in the reduced moments~$R_n$: $am_h$ errors coming from the heavy-quark propagators; $ap$ errors, where $p$ is of order the heavy-quark three-momentum, coming from both quark and gluon propagators; and $a\Lambda_\mathrm{QCD}$ errors, where $\Lambda_\mathrm{QCD}\approx1$~GeV, associated with nonperturbative scales. The $am_h$ errors are seemingly the most concerning since $am_h$ gets as large as 1.21 in our analysis (on the coarsest lattice). \btag{Previous simulations with HISQ quarks} suggest $am_h$~errors here of order
\begin{equation} 
    (m_h/\Lambda_\mathrm{UV})^2=(am_h/\pi)^2\approx15\%
\end{equation}
for $am_h=1$, where $\Lambda_\mathrm{UV}=\pi/a$ is the ultraviolet cutoff on the lattice.\footnote{\btag{One might worry that errors could be of order $(am_h)^2$ (without the $1/\pi^2$). Such errors would be a serious problem, as they would be larger than 100\% when $am_h\ge1$. In fact, however, $am_h$ errors are generally much smaller than this in simulations using HISQ quarks with large $am_h$. See, for example: Figs.~1--4, 7 and~8 in Ref.~\cite{McNeile:2012qf} for various physical quantities involving quarks with $am_h$ as large as~0.85; Figures~1, 4, 5 and~8 in Ref.~\cite{Hatton:2021dvg}, with $am_h$ as large as~0.9; and Figure~2 in Ref.~\cite{Hatton:2021syc}, again with $am_h$ as large as~0.9. In each case $(am_h/\pi)^2$ provides a more realistic estimate of the $am_h$~errors. This choice is also supported by the results presented here in Fig.~\ref{fig:csq}, Table~\ref{tab:Rn}, and Figs.~\ref{fig:Rn_mh} and~\ref{fig:mbmb_a2}.}}
While such errors arise for small~$n$, $am_h$ errors are suppressed by a factor of $(p/m_h)^4\approx(v/c)^4$ as $n$~increases and $p/m_h$~decreases (Fig.~\ref{fig:pv_m}). This factor reduces a 15\% error to around~0.5\% by~$n=14$, at which point $am_h$~errors are negligible relative to the few-percent $(ap/\pi)^2$~and $(a\Lambda_\mathrm{QCD}/\pi)^2$~errors. Thus we expect large, $m_h$-dependent errors at small~$n$ to be rapidly replaced  as $n$~increases by $ap$~and $a\Lambda_\mathrm{QCD}$~errors that are much less dependent on~$m_h$. 

The strong suppression of $am_h$~errors in the nonrelativistic limit is a feature of the HISQ discretization of the quark action used in the simulations. The nonrelativistic behavior 
of any relativistic quark lattice action can be described by a continuum (Euclidean) effective field theory of the form\,\cite{El-Khadra:1996wdx,Lepage:1991ui}
\begin{equation}
    \mathcal{L}_\mathrm{eff} = \psi^\dagger\Big(
    D_t + m_1 - \frac{{\bf D}^2}{2m_2} + \mathcal{O}\big(m(v/c)^4\big)
    \Big) \psi,   
\end{equation} 
where usually $m_1$ and $m_2$ are quite different when $am_h$ is large.\footnote{\btag{For example, tree-level perturbation theory implies $m_1/m_2=0.77$ when $am_2=1$ for Wilson quarks~\cite{El-Khadra:1996wdx}. 
}}
Then the nonrelativistic quark's dispersion relation, relating its energy to its three-momentum~$\pv$ (in lowest-order perturbation theory), 
\begin{equation}
E(\pv)=m_1 + \frac{\pv^2}{2m_2} \big(1+\mathcal{O}(\pv^2/m^2)\big),
\end{equation} 
is wrong. In particular, the quark's ``speed of light'' $c(\pv)$ has the wrong 
limit as $\pv\to0$:
\begin{align}
    \label{eq:csq}
    c^2(\pv) &\equiv \frac{E^2(\pv) - E^2(0)}{{\pv^2}} \\
    &\to \frac{m_1}{m_2}\ne1
\end{align}

The HISQ action is different because it is defined so that the speed of light squared $c^2(\pv)\to1$ as $\pv\to0$ for any value of $am_h$. This implies that $m_1=m_2\equiv m_h^\mathrm{eff}$ for the HISQ action no matter how large $am_h$~is, and therefore HISQ's nonrelativistic effective Lagrangian is exact up to corrections  suppressed by $(v/c)^4$. Consequently the HISQ action is useful for much larger values of $am_h$ than most other discretizations, particularly for nonrelativistic systems.\footnote{In practice the $\epsilon$ parameter in the HISQ action is usually set equal to its value from tree-level perturbation theory~(\eq{eq:epsilon}). As a result $\delta c^2 \equiv c^2(\pv)-1$ is not exactly zero. Consequently there are $a m_h$ errors that are suppressed by $\delta c^2 (v/c)^2$ as well as $(v/c)^4$. This makes little difference since $\delta c^2$ is roughly the same size as or smaller than $(v/c)^2$ for the cases studied here.}

There is no direct way to calculate $c^2(\pv)$ for a heavy quark in simulations, but it is a straightforward calculation for mesons like the~$\eta_b$. In the limit of zero binding, $c^2(\pv)$ would be the same for the meson and quark. Fig.~\ref{fig:csq} shows $c^2(\pv)$ extrapolated to $\pv^2=0$ for $\eta_b$ and $\eta_c$ mesons as a function of the valence-quark mass in lattice units~\cite{[{See Appendix~C in: }] Donald:2012ga}.  The deviations from one are small \btag{(compared with~$(am)^2$)} even when $am>1$; they vanish as $am\to0$. 

\btag{Note that $am_h$ errors are suppressed by only a single power of $(v/c)^2$ in $c^2(\pv\to0)$, because the leading dependence on the rest mass cancels in the numerator of \eq{eq:csq}. As a result $am_h$~errors for the $\eta_b$ in Fig~\ref{fig:csq} should be substantially larger than for the reduced moments~$R_n$ at large~$n$, where errors are suppressed by $(v/c)^4$ and $(v/c)^2$ is small. Note also that the errors in Fig.~\ref{fig:csq} are larger for the $\eta_c$ than for the $\eta_b$ at comparable values of $am_h$, as might be expected since $(v/c)^2$ is larger in the $\eta_c$.}

In Appendix~\ref{app:nrHISQ} we review the derivation of the effective Lagrangian that corresponds to HISQ in the nonrelativistic limit, focusing on its behavior for large values of $am_h$. At leading order in perturbation theory, the theory is characterized by two couplings $c_E(am_h)$ and $c_K(am_h)$ both of which equal one in the continuum limit.
On our finest lattice (EF-5), $am_b=0.62$ and the (tree-level) $am_b$ errors in these couplings are still quite small (less than 13\%): $c_E(am_b)=0.995$ and $c_K(am_b)=0.87$. The $am_b$ corrections are largest for ensemble~SF-5 where $am_b=1.21$. Then $am_b$ errors are still less than 6\% for $c_E$, but $c_K$ has flipped sign: $c_K(am_b)=-0.8$. This leads to errors of relative order $(p/m)^4\approx1\%$ in the $\eta_b$ mass, since $(v/c)^2\approx0.1$ for the valence quarks in an~$\eta_b$.

Another source of $am_h$ errors are the ``algorithmic ghost'' states coming from 
spurious poles in the HISQ propagator (Appendix~\ref{app:ghost}).
Ghost states have energies of order $\pi/(2a)$ and so result in errors of order $(2am_h/\pi)^{n-4}$ for the $n^\mathrm{th}$~moment. These rapidly become insignificant as $n\ge6$ increases.

\subsection{Lattice-spacing uncertainty}
\label{sec:a-uncertainty}
Lattice simulations produce quark and meson masses in lattice units:
for example, heavy-quark masses $am_h$, and the masses $am_{\eta_h}$ for the corresponding pseudoscalar mesons. To be useful, these must be converted to physical units~(GeV), which is done by multiplying by $1/a$. This suggests that the fractional accuracy of lattice results for masses in physical units is limited by the fractional accuracy with which we can determine the lattice spacing (here, about 0.5\%). In fact, however, $b$-quark masses can be far more accurate than the lattice spacing.

We need the lattice $b$ mass~$m_b$ in physical units~(GeV) to determine the $\msb$ mass $\mmsb_b(\mu_n)$ using~\eq{eq:Rnpth}. We obtain $m_b$ by calculating the pseudoscalar-meson masses $am_{\eta_h}$ in simulations for a variety of heavy-quark masses~$am_h$ in the vicinity of the $b$~mass, and then interpolating to find the quark mass that corresponds to the experimental value for the $\eta_b$'s mass~\cite{ParticleDataGroup:2024cfk}:
\begin{equation}
   \label{eq:metab}
    m_{\eta_b}^\mathrm{expt} = 9.3987(20)(10)~\mathrm{GeV},
\end{equation}
where we add a second uncertainty to account for the contribution to~$m_{\eta_b}$ 
from $b\bar b$~annihilation, which is not included in the simulations~\cite{Hatton:2021dvg}. 

The first step in such an analysis is to convert the lattice values
for $am_h$ and $am_{\eta_h}$ to physical units by multiplying by the inverse lattice spacing. This introduces fractional errors into $m_h$ and $m_{\eta_h}$ that are equal to those of~$1/a$, but, for heavy quarks like the~$b$, this uncertainty is strongly suppressed by the interpolation used to obtain the final quark mass~\cite{Chakraborty:2014aca}. 

The suppression occurs because the $\eta_b$~mass is twice the $b$-quark mass plus much smaller binding corrections. This means that $m_{\eta_b}\approx 2m_b$. If the proportionality was exact, such that $m_{\eta_h}=C m_h$ for some constant~$C$, we could determine the $b$-quark mass from lattice data without knowing the lattice spacing: $m_b=m_{\eta_b}^\mathrm{expt}/C$ where $C= (am_{\eta_h})/(am_h)$. Insofar as the proportionality is not exact, dependence on the lattice spacing is suppressed, but not removed (see Appendix~A.1 in~\cite{Chakraborty:2014aca}). As we shall see, the fractional error in $m_b$ is about a fifth that of~$1/a$. Note that the fractional error in $am_b$, therefore, is of order that in the lattice spacing.

The dependence upon the lattice spacing is further reduced by converting the lattice mass to an $\msb$ mass 
using \eq{eq:mbmu1}, where $\rho_n\equiv \mu_n/\mmsb_h(\mu)$ is given in Table~\ref{tab:rn-pth}.
We find that uncertainties coming from the lattice spacing are anti-correlated between the mass~$\mmsb_b(\mu_n)$ and the scale~$\mu_n$, which means that they tend to cancel as the mass is evolved to obtain $\mmsb_b(\mmsb_b)$. As a result of this second level of suppression, the final value for $\mmsb_b(\mmsb_b)$ is largely unaffected by the uncertainty in the lattice spacing.

\subsection{Lattice simulations}
\label{sec:simulations}
The gluon ensembles that we  use are described in Table~\ref{tab:cfg}. They were generated by the MILC collaboration with $n_f=4$ flavors of HISQ sea quark, and have lattice spacings ranging from 0.03--0.06~fm~\cite{Bazavov:2012xda,Bazavov:2017lyh}. The $u$~and $d$~masses are set equal to $m_{\ell}^\mathrm{sea}=(m_{u}^\mathrm{sea}+m_{d}^\mathrm{sea})/2$; corrections to this approximation appear first in second-order chiral perturbation theory (for the quantities we study here) and are negligible \rtag{(less than 0.01\%)}.\footnote{\rtag{The leading-order correction, proportional to $m_\ell$, shifts the $b$~mass (in our analysis) by about~0.06\% according to \eq{eq:fit2} in Sec.~\ref{sec:fit2}. A second-order correction, proportional to $m_u-m_d$, would be at least an order of magnitude smaller.}} 

The $s$ and $c$ sea-quark masses are close to their physical values in all the gluon ensembles.
Three of the four ensembles, however, use $am_{\ell}^\mathrm{sea}\approx m^\mathrm{sea}_s/5$, which is too large. The extent to which the sea-quarks are accurately tuned is measured by the quantities 
\begin{align}
    \delta m_{uds}^\mathrm{sea} &\equiv \sum_{q=u,d,s} (m_q^\mathrm{sea} - m_q) \\
    \delta m_c^\mathrm{sea} &\equiv m_c^\mathrm{sea} - m_c
\end{align}
listed in Table~\ref{tab:cfg}. Here $m_q$ is the correctly tuned lattice mass for quark~$q$. The values given in Table~\ref{tab:cfg} for the $c$-quark mass $m_c$ are from~\cite{Hatton:2021syc}; they were tuned so that the simulations give the correct mass for the $J/\psi$~meson. Here we calculate correctly tuned values for $m_\ell$ and $m_s$ from $m_c$ and the ratios~\cite{FermilabLattice:2018est}
\begin{equation}
   m_c/m_s = 11.771(28) \quad \quad m_s/m_\ell = 27.178(85),
\end{equation}
\rtag{where we have decreased the $m_c/m_s$ ratio by $0.1(1)\%$ from the result in Ref.~\cite{FermilabLattice:2018est} to account for the fact that that reference uses a modified $D_s$~mass to determine the $c$~mass, rather than the $J/\psi$~mass which we use here.\footnote{\rtag{
    Ref.~\cite{FermilabLattice:2018est} uses a scheme where the $c$~mass in the QCD simulation is tuned to reproduce the $D_s$~mass reduced from its experimental value by 1.26~MeV to account approximately for QED corrections. The 1.26~MeV~correction is based on a phenomenological model (see their Eq.~(4.5)). The $c$~mass obtained this way should match that obtained from a similarly QED-corrected $J/\psi$~mass. We do not include QED corrections in our scheme, however, but tune to the experimental value of~$m_{J/\psi}$. A phenomenological model for the $J/\psi$ (tested using lattice QCD+QED~\cite{Hatton:2020lnm}) would increase $m_{J/\psi}/2$ by~1.0~MeV to account for QED binding energy effects (matching~Ref.~\cite{FermilabLattice:2018est}). Therefore our $c$~mass should be about~1~MeV (or~0.1\%) lower than that in Ref.~\cite{FermilabLattice:2018est}. Assigning 100\%~uncertainty to this result, we decrease the $m_c/m_s$ ratio from Ref.~\cite{FermilabLattice:2018est} by $0.1(1)\%$. Corrections of this size have negligible effect~($<0.02\sigma$) on our final result for the $b$~mass.}} 
} 
 
\rtag{Ref.~\cite{FermilabLattice:2018est} also employs a different scheme than we do for tuning the $s$~mass, but the two choices give the same results (within errors). Ref.~\cite{FermilabLattice:2018est} uses a combination of squared $\pi$~and $K$~meson masses from experiment adjusted to allow for QED interaction effects. The combination is chosen to be proportional to~$m_s$ in leading-order chiral perturbation theory. In~\cite{Dowdall:2013rya}, HPQCD determined the mass of a fictitious $s\overline{s}$~meson, known as the~$\eta_s$, which is chosen to be unable to annihilate to gluons. The~$\eta_s$ is a convenient meson to use for tuning the $s$-quark mass once its physical mass in the continuum and chiral limits is known from lattice QCD. Ref.~\cite{Dowdall:2013rya} determined the squared mass of the $\eta_s$ using a ratio to essentially the same combination of squared $\pi$~and $K$~meson masses used 
    in Ref.~\cite{FermilabLattice:2018est}.\footnote{\rtag{The combination used in Ref.~\cite{FermilabLattice:2018est} includes an additional term that they denote by $(M^2_{K^0})^{\gamma}$ but this is very small and easily covered by that fact that the HPQCD combination includes a much larger uncertainty on the possible violations of Dashen's theorem~\cite{Dowdall:2013rya}.}} 
The ratio is expected to be close to~1 from chiral perturbation theory and that is indeed found to be the case. Using this ratio, HPQCD determined $m_{\eta_s}= 688.5(2.2)~\mathrm{MeV}$~\cite{Dowdall:2013rya} in a pure lattice QCD calculation. Tuning~$m_s$ using this value is then equivalent to the strategy adopted in Ref.~\cite{FermilabLattice:2018est}.}

\rtag{As discussed in the previous section, we tune the $b$-quark mass so that simulations give the experimental $\eta_b$~mass (\eq{eq:metab}). Again our scheme differs slighlty from that used in Ref.~\cite{FermilabLattice:2018est}, where the $b$~mass is tuned to reproduce a modified value of the $B_s$~mass. This difference will be relevant in Sec.~\ref{sec:smass} where we use the mass ratio $m_b/m_s$ from Ref.~\cite{FermilabLattice:2018est}. As above for $m_c/m_s$, we can use phenomenological models~\cite{FermilabLattice:2018est,Gregory} to show that our $b$~mass should be about 0.8~MeV (or 0.02\%) lower than that from Ref.~\cite{FermilabLattice:2018est}. Taking the correction as~$-0.02(2)\%$, the mass ratio becomes 
\begin{equation}
    \label{eq:mb_ms}
    m_b/m_s = 53.93(12),
\end{equation}
which is an inconsequential shift from the original value of~$53.94(12)$.}
 
\rtag{Finally, we tune the $\ell$-quark mass to reproduce the experimental mass of the $\pi^0$~meson, which is consistent with what is done in Ref.~\cite{FermilabLattice:2018est}.}
Increasing the light-quark mass $m_{\ell}^\mathrm{sea}$ in the sea makes simulations much less costly because smaller lattice volumes can be used (without increasing finite-volume errors). The simulations are then unphysical, but they are easily corrected~\cite{Chakraborty:2014aca}. Ensemble~SF-P uses sea quarks that are all close to their physical values. 

The finest lattices (UF-5 and~EF-5) show signs of long Monte Carlo correlation times in measurements of the net topological charge. A detailed analysis using chiral perturbation theory, tested against lattice results, shows that visible effects from this topological freezing are limited to mesons with $u/d$~valence quarks; the impact on simulations of the $D_s$~meson, for example, is negligible~\cite{Bernard:2017npd}. This seems reasonable because only very light quarks can sample topology over the entire lattice volume. Effects will be much smaller still for systems like the $\eta_b$ where both valence quarks are heavy.

\begin{table*}
    \caption{Gluon ensembles used in this paper. These were produced by the MILC collaboration~\cite{Bazavov:2012xda,Bazavov:2017lyh}. Lattice spacings are determined from the Wilson flow parameter $w_0/a$ measured for each ensemble, assuming $w_0=0.1715(9)$~fm~\cite{Dowdall:2013rya}. The spatial ($L/a$) and temporal ($T/a$) sizes of the lattice are listed, as are the sea-quark masses for $\ell$, $s$, and $c$~quarks used in the simulation, where $u$~and $d$~masses are set equal to their average~$m_{\ell}$. $m_c$ is the valence $c$~mass (in GeV) needed so that the $J/\psi$~meson mass obtained in the simulation agrees with experiment; these values come from an earlier analysis~\cite{Hatton:2021syc} using data from~\cite{Hatton:2020qhk}. Similarly $m_b$ (in GeV) is tuned to give the correct value for the $\eta_b$~mass. The last two entries indicate the extent to which the sea-quark masses are detuned in the simulations. 
    SF, UF, and EF are abbreviations for ``superfine,'' ``ultrafine,'', and ``exafine.''
    Ensembles SF-5, SF-P, UF-5, and EF-5 have lattice spacings of 0.059, 0.057, 0.044, and 0.032~fm, respectively.}
    \label{tab:cfg}
    \begin{ruledtabular}
        \begin{tabular}{clccrccccccc}
            Ensemble & $w_0/a$ & $L/a$ & $T/a$ & $N_\mathrm{cfg}$ 
            & $am_{\ell}^\mathrm{sea}$ & $am_{s}^\mathrm{sea}$ 
            & $am_{c}^\mathrm{sea}$ & $m_c$ & $m_b$ 
            & $\delta m_{uds}^\mathrm{sea}/m_s$ 
            & $\delta m_c^\mathrm{sea}/m_c$ \\ 
            \hline
            SF-5 & 2.8941(46) & 48 & 144 & 1017 & 0.00480 & 0.0240 & 0.2859 & 0.912(3) & 4.0204(43) & \rtag{$0.371(10)\quad\quad$} & \rtag{$0.044(6)\quad$} \\
            SF-P & 3.0170(23) & 96 & 192 &  100 & 0.00080 & 0.0220 & 0.2600 & 0.897(3) & 3.9900(41) & \rtag{$0.002(7)\quad\quad$ } & \rtag{$0.007(6)\quad$} \\
            UF-5 & 3.892(12)  & 64 & 192 &  200 & 0.00316 & 0.0158 & 0.1880 & 0.868(3) & 3.8739(47) & \rtag{$0.270(10)\quad\quad$} & \rtag{$-0.030(7)\quad$} \\
            EF-5 & 5.243(16)  & 96 & 288 &  100 & 0.00223 & 0.0111 & 0.1316 & 0.825(3) & 3.7332(49) & \rtag{$0.270(10)\quad\quad$} & \rtag{$-0.038(7)\quad$} \\
        \end{tabular}
    \end{ruledtabular}
\end{table*}

We determine the lattice spacing for each gluon ensemble by measuring the Wilson Flow parameter $w_0/a$ in the simulations~\cite{Borsanyi:2012zs}, where from previous simulations~\cite{Dowdall:2013rya},
\begin{equation}
    w_0 = 0.1715(9)\,\mathrm{fm}.
\end{equation}

Table~\ref{tab:Rn} contains our lattice results for the pseudoscalar $\eta_h$ meson mass and for the reduced moments~$R_n$ evaluated for a variety of heavy-quark masses~$am_{h}$ using each of the four gluon ensembles. We restrict our analysis to heavy-quark masses that are larger than $0.65m_b$ so that the perturbation theory for the $r_n$ (\eq{eq:rn-pth}) converges rapidly.

The only source of uncertainty in $R_n$ comes from statistical Monte Carlo errors in the moments $G_n$ of the correlators. These uncertainties are extremely small (a part in $10^7$ or $10^8$) for the pseudoscalar correlators we are using. Fitting data to this level of precision would require a very complicated model. To simplify our analysis, we increase the uncertainty in each reduced moment by multiplying by $1 \pm \sigma_u$, where $\sigma_u$ represents the part of the variation in the data that is \emph{unexplained} by our model~\cite{Hatton:2021syc}. We use the same $\sigma_u$ for each $R_n$, and the corresponding uncertainties are uncorrelated from one $R_n$ to another. We  adjust $\sigma_u$ to maximize the Bayes Factor for the fit (while requiring a good fit)~\cite{[{See Sec.~5.2 on the Empirical Bayes method in }] Lepage:2001ym}. For the fit model we use here,  the optimal~$\sigma_u$ is
\begin{equation}
    \label{eq:sigma_u}
    \sigma_u = 0.0001.
\end{equation}
Using 0.00005 or 0.0002 instead of 0.0001 has \rtag{negligible impact~($<0.1\sigma$)} on our final results. A simpler model would require a larger~$\sigma_u$.

The $\eta_h$ masses are obtained by fitting the pseudoscalar correlator to a multiexponential formula for the correlator:
\begin{equation}
    \sum_{j=1}^2 a_j \big(\mathrm{e}^{-E_j t} + \mathrm{e}^{-E_j(T-t)}\big).
\end{equation}
We restrict the fit to the 9~$t$s centered on~$T/2$ for each lattice, to minimize the impact of excited states.\footnote{\rtag{The first excited state of the $\eta_b$, the $\eta_b(2\mathrm S)$, contributes less than~0.0004\% to the $\eta_b$~correlator over the $t$-range used in the fits. 
This makes a negligible contribution to the 0.1\%~uncertainty in the $b$-quark masses in Table II, which are derived from the $\eta_b$~masses.  $T$~ranges from 8.5~to 10.9~fm for the different configuration sets.}} This is feasible for pseudoscalar correlators because the Monte Carlo signal-to-noise ratio remains constant as $t$~increases for these correlators. Using two or four times as many $t$s gives the same results. Adding more terms to the sum also has no impact on the results \rtag{(to the precision shown in Table~\ref{tab:Rn})}, and neither does restricting the sum to only one term. The results also agree well with simple effective-mass determinations.\footnote{\rtag{Fit results are compared with effective-mass determinations of the $\eta_h$~mass in Fig.~1 of~\cite{Hatton:2021syc}. These are for the largest mass used on the EF-5~configurations.}}

The $b$-quark mass $m_b$ in Table~\ref{tab:cfg} is chosen so that the corresponding $\eta_b$ mass in the simulation agrees with experiment~(\eq{eq:metab}). The highest heavy-quark mass entries for gluon ensembles~SF-5 and~SF-P, and the next-to-highest masses for ensembles~UF-5 and~EF-5 in Table~\ref{tab:Rn} are very close to the correct $b$~masses for those ensembles. We refine these estimates of $m_b$ by fitting the heavy-quark masses $a m_h$ for each ensemble as a function of the $a\eta_h$ masses with a monotonic spline~\cite{Steffen:1990,[{The splines are implemented using the \texttt{gvar} Python module: }] peter_lepage_gvar}. We then interpolate between the heavy-quark masses with the spline to obtain~$am_b$. 

The difference between the interpolated value and the nearest simulated value is negligible (0.01\%--0.3\%), but the interpolation itself suppresses the dependence of our results on the uncertainties in the lattice spacings (Sec.~\ref{sec:a-uncertainty}). As a result uncertainties in the $m_b$s in Table~\ref{tab:cfg} are around 0.1\%, while the uncertainties in the lattice spacings (that is, from $w_0/a$ and $w_0$) are much larger~(0.5\%). There is a similar but smaller suppression for the $c$-quark masses~$m_c$ in Table~\ref{tab:cfg}.

\begin{table*}
    \caption{Simulation results for the $\eta_h$ mass $am_{\eta_h}$ and for the reduced moments $R_n$ defined in \eq{eq:Rn}. Results are given for different valence heavy-quark masses $am_{h}$ and for different gluon ensembles (see Table~\ref{tab:cfg}). $am_{0h}$ is the mass parameter, used in  the HISQ action, corresponding to $am_h$ (\eq{m0eqn}). Statistical errors (and correlations) in the results for $am_{\eta_h}$ and $R_n$ are smaller than~1 in the last decimal place shown,  and have negligible impact on our final $b$~mass~\rtag{($<0.002\sigma$)}. \rtag{The $\eta_h$ masses were recalculated for this paper but agree with the results given in~\cite{Hatton:2021syc}.}}
    \label{tab:Rn}
    \begin{ruledtabular}
        \begin{tabular}{ccccccccccccc}
            Ens. & $am_{0h}$ & $am_h$ & $am_{\eta_h}$ & $R_6$ & $R_8$ & $R_{10}$ & $R_{12}$ & $R_{14}$ & $R_{16}$ & $R_{18}$ & $R_{20}$ & $R_{22}$ \\ \hline
            SF-5 & 0.800 & 0.7901 & 1.9875 & 1.0281 & 0.9924 & 0.9697 & 0.9515 & 0.9367 & 0.9249 & 0.9152 & 0.9072 & 0.9005 \\
            & 0.900 & 0.8832 & 2.1758 & 1.0300 & 1.0018 & 0.9837 & 0.9683 & 0.9548 & 0.9434 & 0.9338 & 0.9258 & 0.9190 \\
            & 1.000 & 0.9735 & 2.3577 & 1.0302 & 1.0077 & 0.9938 & 0.9813 & 0.9698 & 0.9593 & 0.9501 & 0.9421 & 0.9353 \\
            & 1.280 & 1.2086 & 2.8251 & 1.0276 & 1.0146 & 1.0081 & 1.0024 & 0.9965 & 0.9903 & 0.9840 & 0.9778 & 0.9719 \\
            SF-P & 0.760 & 0.7521 & 1.9093 & 1.0267 & 0.9876 & 0.9630 & 0.9438 & 0.9288 & 0.9168 & 0.9072 & 0.8993 & 0.8927 \\
            & 0.800 & 0.7901 & 1.9863 & 1.0281 & 0.9925 & 0.9698 & 0.9516 & 0.9368 & 0.9250 & 0.9154 & 0.9074 & 0.9007 \\
            & 1.000 & 0.9735 & 2.3567 & 1.0302 & 1.0077 & 0.9938 & 0.9813 & 0.9698 & 0.9593 & 0.9501 & 0.9422 & 0.9353 \\
            & 1.210 & 1.1525 & 2.7137 & 1.0284 & 1.0137 & 1.0059 & 0.9988 & 0.9916 & 0.9843 & 0.9771 & 0.9703 & 0.9640 \\
            UF-5 & 0.600 & 0.5974 & 1.5491 & 1.0103 & 0.9615 & 0.9333 & 0.9140 & 0.9000 & 0.8893 & 0.8808 & 0.8738 & 0.8680 \\
            & 0.800 & 0.7901 & 1.9458 & 1.0232 & 0.9915 & 0.9710 & 0.9545 & 0.9412 & 0.9306 & 0.9219 & 0.9148 & 0.9089 \\
            & 0.880 & 0.8648 & 2.0982 & 1.0248 & 0.9985 & 0.9814 & 0.9669 & 0.9544 & 0.9440 & 0.9353 & 0.9281 & 0.9220 \\
            & 0.900 & 0.8832 & 2.1356 & 1.0250 & 0.9999 & 0.9836 & 0.9696 & 0.9574 & 0.9471 & 0.9384 & 0.9312 & 0.9251 \\
            EF-5 & 0.403 & 0.4026 & 1.1009 & 0.9748 & 0.9149 & 0.8855 & 0.8671 & 0.8540 & 0.8441 & 0.8361 & 0.8295 & 0.8239 \\
            & 0.450 & 0.4493 & 1.2013 & 0.9852 & 0.9289 & 0.9003 & 0.8822 & 0.8693 & 0.8595 & 0.8516 & 0.8452 & 0.8397 \\
            & 0.550 & 0.5483 & 1.4107 & 1.0018 & 0.9535 & 0.9267 & 0.9090 & 0.8963 & 0.8867 & 0.8790 & 0.8727 & 0.8674 \\
            & 0.622 & 0.6189 & 1.5581 & 1.0097 & 0.9675 & 0.9426 & 0.9252 & 0.9126 & 0.9029 & 0.8953 & 0.8890 & 0.8837 \\
            & 0.650 & 0.6462 & 1.6149 & 1.0120 & 0.9721 & 0.9480 & 0.9309 & 0.9183 & 0.9086 & 0.9010 & 0.8947 & 0.8894 \\    
        \end{tabular}
    \end{ruledtabular}
\end{table*}

\subsection{Fitting lattice values for $R_n$}
\label{sec:Rn-fit}
The largest systematic errors that arise when using \eq{eq:Rnpth} to determine the $b$-quark's $\msb$ mass $\mmsb_b(\mu_n)$ come from finite-lattice-spacing errors in the lattice determinations of the reduced moments~$R_n$ and the bare lattice quark masses~$m_b$ for each gluon ensemble~$s$. It is convenient to address the systematic errors in $R_n$ and $m_b$ separately, in two steps:
\begin{enumerate}
    \item We first fit the~$R_n$ from each gluon ensemble~$s$ to determine an $\msb$ $b$-quark mass $\mmsb_b^{(s)}(\mmsb_b^{(s)})$ for each ensemble separately. In doing so we correct for systematic errors in the lattice values for the $R_n$ and in the perturbation theory for the~$r_n$. We use the $m_b$ values given in Table~\ref{tab:cfg}.

    \item The four $\mmsb_b^{(s)}(\mmsb_b^{(s)})$ obtained in the first fit do not agree with each other because of systematic errors in the lattice results for the $\eta_b$~mass used to determine the bare $b$-quark mass $m_b$ for each gluon ensemble. These are corrected in a second fit which extrapolates the $\msb$ masses to zero lattice spacing, while also accounting for the detuned sea-quark masses used to create the different ensembles. 
\end{enumerate}
Here we discuss the first step on this analysis; the second step is discussed in the Sec.~\ref{sec:fit2}.

We fit results for the reduced moments from each gluon ensemble~$s$ using the following formula:
\begin{align}
    \label{eq:fit1}
    R_n^{(s)}(m_h) &= \frac{m_b}{\mmsb_b^{(s)}(\mu_n)} \,
    r_n\big(\almsb(\xi_\alpha \mu_n), \rho_n\big)
    \nonumber \\
    &\times \Bigg(
        1 + \sum_{j=1}^{N_{a}} c_j^{(n,s)}(m_h/m_b)\,
        \Big(\frac{a\Lambda_n(m_h)}{\pi}\Big)^{2j}
        \Bigg) \nonumber \\
    &\times \big(1 + d_n^\mathrm{pth} \rtag{\,r_{n2}\,\almsb^2(\mu_n) \, (m_c/\mu_n)^2}\big),
\end{align}
where 
\begin{equation}
    \mu_n = \rho_n \frac{am_h}{am_b}\,\mmsb_b^{(s)}(\mu_n)
\end{equation}
and $\rho_n$ is specified in Table~\ref{tab:rn-pth}.
We fit $R_n^{(s)}(m_h)$~values for every ensemble~$s$, heavy-quark mass~$m_h$, and moment~$n$ listed in Table~\ref{tab:Rn}. These fits are all done together because some fit parameters are shared across all the fits. We use a Bayesian fit with priors for every fit parameter~\cite{Lepage:2001ym,[{Fits were done using the \texttt{lsqfit} Python module: }] peter_lepage_lsqfit}. The priors are \emph{a priori} estimates of the parameters based on earlier analyses and/or theoretical estimates. \rtag{Quantities in \eq{eq:fit1} whose values are known from earlier analyses but have uncertainties, like $m_b$ (from Table~\ref{tab:cfg}) and $\almsb$ (see below), are treated as fit parameters whose priors are set equal to the earlier results.}

We now examine each component of the fit in turn.

\subsubsection{Truncated perturbation theory}
We truncate the perturbative expansion for $r_n$ (\eq{eq:rn-pth}):
\begin{equation}
    \label{eq:rn-tpth}
   r_n
   = 1 + \sum_{j=1}^{N_\mathrm{r}} r_{nj}(\rho_n) \,\almsb^j(\xi_\alpha\mu_n).
\end{equation}
The coefficients for the first three orders are given in Table~\ref{tab:rn-pth}; coefficients for higher-order terms are set equal to $0\pm r_n^\mathrm{max}$ where $r_n^\mathrm{max}$ is given by \eq{eq:rnmax}. We take $N_\mathrm{r}=4$, to allow for higher-order corrections, but we get the same final results with any $N_\mathrm{r}\ge3$. Coefficients with uncertainties become fit parameters whose priors equal their values in Table~\ref{tab:rn-pth} (or $0\pm r_n^\mathrm{max}$). Doubling $r_n^\mathrm{max}$ has \rtag{negligible effect ($<0.02\sigma$) on our final $b$~mass.} 
\rtag{The Empirical Bayes method~\cite{[{See Sec.~5.2 on the Empirical Bayes method in }] Lepage:2001ym} indicates that a value slightly less than~$r_n^\mathrm{max}/2$ is favored by the fit data, but we choose to continue with $r_n^\mathrm{max}$, which is more conservative.}

The factor $\xi_\alpha$ in the $\msb$ coupling $\almsb(\xi_\alpha\mu_n)$ corrects for any detuning of the sea quark masses used to generate the gluon ensembles. It is given by~\cite{Chakraborty:2014aca}
\begin{align}
    \xi_\alpha = 1 &+ g_\alpha \frac{\delta m_{uds}^\mathrm{sea}}{m_s} 
    + g_{a^2,\alpha} \frac{\delta m_{uds}^\mathrm{sea}}{m_s} \Big(\frac{m_c}{\pi/a}\Big)^2 
    \nonumber \\
    &+ g_{c,\alpha} \frac{\delta m_c}{m_c} + \mathcal{O}(\delta m^2)
    \label{eq:xi_alpha}
\end{align}
where $\delta m^\mathrm{sea}_{uds}$ and $\delta m^\mathrm{sea}_c$ are listed in Table~\ref{tab:cfg}, and fit parameters $g_\alpha$, $g_{a^2,\alpha}$, and $g_{c,\alpha}$ have the following priors~\cite{Chakraborty:2014aca}:
\begin{equation}
    g_\alpha = 0.082(8)\quad g_{a^2,\alpha} = 0.0(1) \quad g_{c,\alpha} = 0.0(1).
\end{equation}

We parameterize the coupling $\almsb(\mu)$ by  its value at 5~GeV, taking 
$\almsb(5)$ as a fit parameter with prior~\cite{Chakraborty:2014aca}
\begin{equation}
    \label{eq:al5}
    \almsb(5~\mathrm{GeV}) = 0.2128(25)
\end{equation}
for $n_f=4$ sea-quark flavors.
The coupling at other values of $\mu$ is obtained by integrating (numerically) the evolution equation
\begin{align}
    \label{eq:evol-alpha}
    \mu^2\frac{d\almsb(\mu)}{d\mu^2} = -  \almsb^2(\mu) \sum_{j=0}^{4}\beta_j(n_f) \,\almsb^j(\mu).
\end{align}
Using a prior of $0.21(1)$ instead of $0.2128(25)$ for $\almsb(5)$ has negligible effect~\rtag{($<0.02\sigma$)} on our final results for the $b$-quark mass. Taking $\almsb(5)=0.2128$, with no uncertainty, also has negligible effect~\rtag{($<0.02\sigma$)}.\footnote{Our final result for the $b$~mass is quite insensitive to the value of the coupling: it changes by $1\sigma$ or less as the value of $\almsb(M_Z,n_f=5)$ is varied from 0.1148 to~0.1206. The coupling used in this paper (\eq{eq:al5}) corresponds to $\almsb(M_Z,n_f=5)=0.1182(7)$~\cite{Chakraborty:2014aca}.}

We implement the $\msb$ mass $\mmsb_b^{(s)}(\mu_n)$ in \eq{eq:fit1} by introducing a fit parameter, with (noninformative) prior $4.2(1.0)$, corresponding to 
the mass's value at scale
\begin{equation}
    \label{eq:mu-s}
    \mu^{(s)}=\mmsb_b^{(s)}(\mu^{(s)}).
\end{equation}
Again we integrate its evolution equation numerically to obtain masses for other values of $\mu$:
\begin{equation}
    \label{eq:evol-m}
    \mu^2\frac{d\log(\mmsb_b^{(s)}(\mu))}{d\mu^2} = -\almsb(\mu) \sum_{j=0}^4 \gamma_j(n_f)\, \almsb^j(\mu),
\end{equation}
where~$n_f=4$.
Masses $\mmsb_h^{(s)}(\mu)$ for other heavy quarks are related to $\mmsb_b^{(s)}(\mu)$ by (\eq{eq:mmsb_m})
\begin{equation}
    \mmsb_h^{(s)}(\mu) = \frac{am_h}{am_b} \, \mmsb_b^{(s)}(\mu),
\end{equation}
where $m_h$ and $m_b$ are the corresponding lattice masses for gluon ensemble~$s$ from Tables~\ref{tab:Rn} and~\ref{tab:cfg}, respectively.

The final output values for the four fit parameters (\eq{eq:mu-s}) are the input data used in the second stage of our analysis, described in Sec.~\ref{sec:fit2}.

The perturbative coefficients used in the evolution equation are from~\cite{vanRitbergen:1997va,Czakon:2004bu,Vermaseren:1997fq,Chetyrkin:1997dh,Chetyrkin:1997un,Herzog:2017ohr,Baikov:2014qja}. The software used to integrate the equations numerically is from~\cite{peter_lepage_qcdevol}. Adding an additional term (for $j=5$) in the evolution equations Eqs.~(\ref{eq:evol-alpha}) and~(\ref{eq:evol-m}) has negligible effect~\rtag{($<0.01\sigma$)} on our final results.

\subsubsection{$a\Lambda_n$ errors}
The largest errors in the lattice values for $R_n^{(s)}$ are from the finite lattice spacings used in the simulations. As discussed in Sec.~\ref{sec:finite-a}, these are dominated by errors of order $(ap/\pi)^{2j}$ where $p$ is the typical three momentum of the valence heavy-quarks. We correct for this error by including a factor 
\begin{equation}
    \label{eq:aLamn}
    1 + \sum_{j=1}^{N_{a}} c_j^{(n,s)}(m_h/m_b)\,
    \Big(\frac{a\Lambda_n(m_h)}{\pi}\Big)^{2j}
\end{equation}
in the fit, where 
\begin{equation}
    \label{eq:Lam_n}
    \Lambda_n(m_h) \approx p \approx m_h\,\frac{3.15}{n^{0.75}}
\end{equation}
based on the results shown in Fig.~\ref{fig:pv_m}. 

We write the coefficients as a sum of two functions
\begin{equation}
    c_j^{(n,s)}(m_h/m_b) \equiv c_{0j}^{(n)}(m_h/m_b) + \delta c_{j}^{(n,s)}(m_h/m_b).
\end{equation}
The first  function $c_{0j}^{(n)}(m_h/m_b)$ is common to all gluon ensembles. The second function allows for non-analytic dependence on the lattice spacing (for example, $\almsb(\pi/a)$ corrections) and also such things as small differences in the sea-quark masses; it can differ from gluon ensemble to ensemble. These functions are implemented as monotonic splines~\cite{peter_lepage_gvar} with knots at
\begin{equation}
    x_k \in \{0.65, 0.85, 1.05\}.
\end{equation}
where $x\equiv m_h/m_b$.
The values of the functions at the knots are fit parameters with (uncorrelated) priors 
\begin{align}
    c_{0j}^{(n)}(x_k) &= 0(1) \\
    \delta c_j^{(n,s)}(x_k) &= 0.0(5).
\end{align}
The Empirical Bayes method~\rtag{\cite{[{See Sec.~5.2 on the Empirical Bayes method in }] Lepage:2001ym}} suggests priors that are half as wide as these, which results in a good fit, \rtag{a mass that is shifted up by $\sigma/3$,} and an uncertainty that is reduced by~$0.16\sigma$. We will use the wider prior here, to be conservative.\footnote{\rtag{Doubling the widths of the priors increases the $b$-quark mass's uncertainty by $0.14\sigma$ and decreases the fit's $\chi^2$ by~40\%, but also decreases the Bayes Factor by more than thirty orders of magnitude~($9\times10^{-32}$)\,---\,the fit data strongly prefers the narrower prior. Note that fit values for the coefficients are $\mathcal{O}(1)$ or smaller in magnitude, even for the wider prior (width~2): the fit value for the largest coefficient (in magnitude) is~$-1.09(6)$ for the original prior and $-1.09(9)$ with the wider prior.
The largest coefficients are for $n=6$ moments; the largest (in magnitude) coefficient for~$n=22$ 
is~$-0.53(15)$.}}

The knots are uniformly distributed across the range of values for $m_h/m_b$ in our analysis. Fitting with just two knots does not give good fits ($\chi^2$ too large). Fitting with four knots gives the same results for the $b$~mass as from three knots, but with a much smaller Bayes Factor. 

We include only $a^2$ and $a^4$ corrections ($N_{a}=2$). Adding further terms has negligible effect~($<0.02\sigma$) on our final results.

\subsubsection{Nonzero $m_c$}
As mentioned above, the perturbative coefficients $r_{n2}$ and $r_{n3}$ in Table~\ref{tab:rn-pth} are calculated assuming zero sea-quark masses. While this approximation is reasonable for $u$, $d$, and $s$~quarks, there could be appreciable corrections, of order $r_{n2}\,\almsb^2(\mu_n)$ times $(m_c/\mu_n)^2$, for the $c$~quark. \rtag{We allow for such corrections by including a factor 
\begin{equation}
    \label{eq:nonzero-mc}
    1 + d^\mathrm{pth}_n \,r_{n2}\,\almsb^2(\mu_n) \,\big(m_c/\mu_n\big)^2
\end{equation}
in the fit formula, where fit parameter $d^\mathrm{pth}_n$ has prior~$0(1)$. The Empirical Bayes method~\cite{[{See Sec.~5.2 on the Empirical Bayes method in }] Lepage:2001ym} indicates a prior that is three quarters as wide, but changing to this prior would have negligible impact~($<0.02\sigma$) on the final result. Doubling the prior width also has negligible effect ($<0.02\sigma$).}

\subsubsection{Nonperturbative and finite-volume effects}
There are nonperturbative corrections to perturbative expression for the reduced moments (\eq{eq:Rnpth}) coming from gluon and light-quark condensates, but these are suppressed by a factor of $1/m_h^4$. Condensates were negligible (less than 0.05\%) in our earlier determination of the $c$-quark mass~\cite{Chakraborty:2014aca}; they should be much smaller here since the quark masses are 2--3~times larger. We verified this by repeating the present analysis with and without condensates; they make no difference to the final results at our level of precision. 

Finite-volume errors can enter through the condensates, but again these are negligible in the $c$-mass analysis (less than 0.01\%) and would be much smaller here because of the larger quark masses.

\subsection{Extrapolating $\mmsb_b(\mmsb_b)$ to the continuum limit}
\label{sec:fit2}
The analysis above generates a $b$~mass 
$\mmsb_b^{(s)}(\mmsb_b^{(s)})$ from each gluon ensemble~$s$ (\eq{eq:mu-s}).
The second stage of our analysis corrects these results for detuned sea-quark masses, and extrapolates them to zero lattice spacing, thereby correcting for finite-lattice-spacing errors in the $\eta_b$ masses used to determine $m_b$ for each gluon ensemble. We do this by fitting $b$-mass results for each ensemble~$s$ with the following formula:
\begin{align}
    \mmsb_b^{(s)}(\mmsb_b^{(s)}) = \,\, &\xi_m \mmsb_b(\xi_\alpha \mu)
    \nonumber \\
     &\times  
    \Big(
        1 + \sum_{j=1}^{N_{a}} f_j \Big(\frac{a\Lambda_{\eta_b}}{\pi}\Big)^{2j}
    \Big)
    \label{eq:fit2}
\end{align}
where $\mu\equiv\xi_m \mmsb_b(\xi_\alpha \mu)$, $\mmsb_b(\mu)$ is the physical (continuum) $b$-quark's $\msb$ mass, and $\Lambda_{\eta_b}\approx 0.3\,m_b$ is of order the valence quarks' momentum in an~$\eta_b$ (Sec.~\ref{sec:finite-a}). We again take $N_a=2$; results are unchanged for any~$N_a\ge2$. \rtag{The $f_j$ fit parameters have prior $0(1)$. The Empirical Bayes method indicates a prior width 1.5~times wider than what we use, but this has negligible effect~($<0.02\sigma$) on the final uncertainty of the $b$~mass and shifts the mass up by less than $0.1\sigma$.}

We implement the $\msb$~mass $\mmsb_b(\mu)$ in this formula by again introducing a fit parameter corresponding to the mass's value  at scale
\begin{equation}
    \mu = \mmsb_b(\mu)
\end{equation} 
The mass for other scales is obtained by integrating the quark-mass evolution equation~(\eq{eq:evol-m}) numerically. The best-fit output value for the fit parameter is our final result for the 
$n_f=4$ $b$-quark mass $\mmsb_b(\mmsb_b)$ (without QED).

The $\xi_m$ and $\xi_\alpha$ factors in $\xi_m \mmsb_b(\xi_\alpha \mu)$ account for the detuned sea-quark masses~\cite{Chakraborty:2014aca}. $\xi_\alpha$ is discussed above (\eq{eq:xi_alpha}). Again $g_\alpha$, $g_{a^2,\alpha}$, and $g_{c,a}$ are fit parameters. $\xi_m$ is given by (for $b$ quarks)
\begin{align}
    \xi_m = 1 &+ \tilde g_m \frac{\delta m_{uds}^\mathrm{sea}}{m_s} 
    \nonumber \\
    &+ \tilde g_{a^2,m} \frac{\delta m_{uds}^\mathrm{sea}}{m_s}
    \Big(\frac{am_c}{\pi}\Big)^2
    \label{eq:xi_m}
\end{align}
where fit parameters $\tilde g_m$ and $\tilde g_{a^2m}$ have priors~\cite{Chakraborty:2014aca}
\begin{equation}
    \tilde g_m = 0.0248(45)\quad \tilde g_{a^2,m} = 0.0(1).
\end{equation}

\section{Lattice QCD Results (Without QED)}
\label{sec:qcd-results}

\begin{figure}
    \begin{center} 
    \includegraphics[scale=0.8]{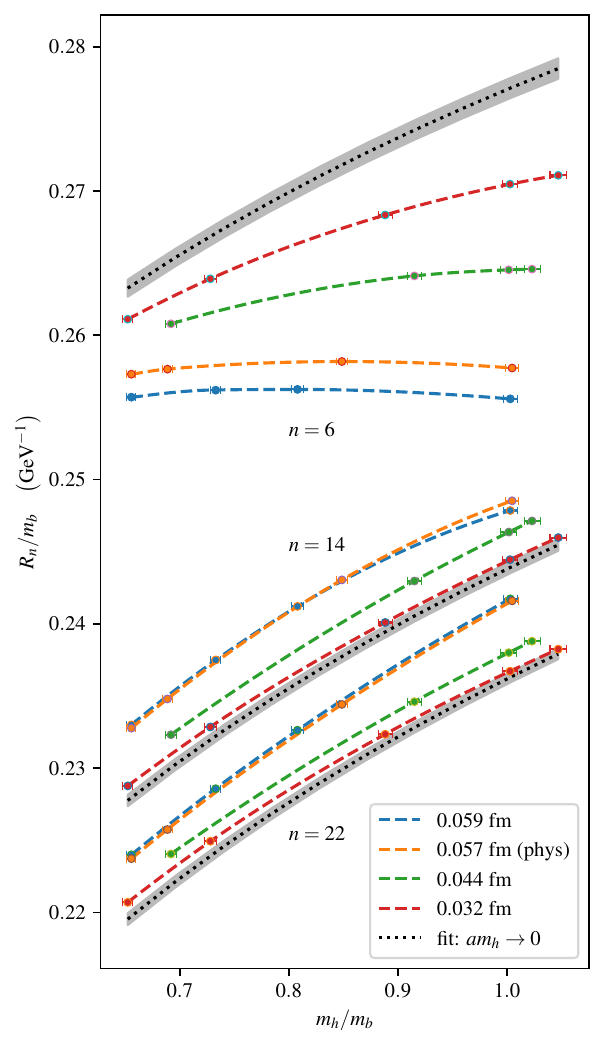}    
    \end{center}
\caption{\label{fig:Rn_mh} Fits to lattice values for the reduced moments~$R_n$. Results for $R_n/m_b$ are plotted for a range of (large) valence-quark masses~$m_h$. The data points are lattice results for moments with $n=6$, 14, and~22. These are calculated using gluon ensembles~SF-5~(blue), SF-P~(orange), UF-5~(green), and EF-5~(red), with lattice spacings of 0.059, 0.057, 0.044, and 0.032~fm, respectively. The lattice results are compared with fit results (dashed lines) based on \eq{eq:fit1}. The dotted grey lines and 1-sigma grey bands are fit results with $am_h$ set to zero.}
\end{figure}

Fig.~\ref{fig:Rn_mh} shows results, as a function of $m_h/m_b$, from our fit of \eq{eq:fit1} to lattice values for the reduced moments~$R_n$ with $n=6$, 14, and~22. (Plots for the other moments are similar.) We plot $R_n/m_b$ because that quantity would be independent of the lattice spacing if there were no finite lattice-spacing errors. Fit results (dashed lines) are shown with the lattice results for each of the gluon ensembles. We also show the fit results extrapolated to $am_h=0$ for each reduced moment (dotted lines and grey bands). 

\rtag{The fit has a $\chi^2$ per degree of freedom of~0.18 with 153 degrees of freedom ($p=1$).} The $\chi^2$ is low because the priors used in the fit are generally conservative. In such situations it is important to add noise to the priors in order to test for goodness of fit~\cite{[{See Appendix~D.4 in: }] Dowdall:2019bea}.  With noise, the $\chi^2$ per degree of freedom typically ranges between~0.9 and~1.1, as expected for a good fit to this much data. 

The finite lattice-spacing errors for the $n=6$ moments are strikingly different from those for $n=14$ and~22. 
The $n=6$ moments show large errors with strong $m_h$~dependence. 
These errors become much smaller and much less dependent on~$m_h$ as $n$~increases. This is because the large~$n$ moments are nonrelativistic (Fig.~\ref{fig:pv_m}). As discussed in Sec.~\ref{sec:finite-a}, $am_h$ errors are suppressed in nonrelativistic quantities (by $(v/c)^4$), allowing other finite-lattice-spacing errors to become competititve. Here results from the smallest lattice spacing (EF-5, 0.032~fm) are very close to the final fit results for $am_h=0$ when~$n\ge14$, particularly around~$m_h=m_b$.

\begin{figure}
    \begin{center}
    \includegraphics[scale=0.9]{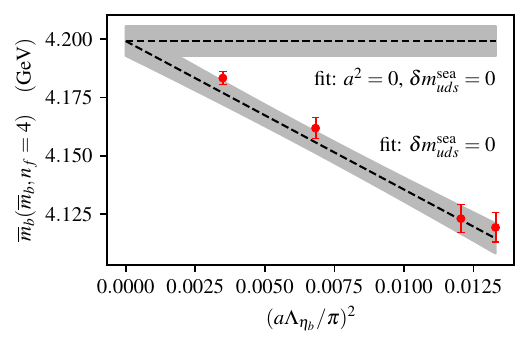}    
    \end{center}
\caption{\label{fig:mbmb_a2} Fit to results for the $b$-quark masses $\mmsb_b^{(s)}(\mmsb_b^{(s)})$ from each gluon ensemble~$s$. Fit results (dashed line and grey bar), using \eq{eq:fit2}, are shown for tuned sea quarks ($\delta m_{uds}^\mathrm{sea}\to0$) and for zero lattice spacing. 
The first, second, and fourth masses are from simulations where the light sea-quark mass $m_\ell$ was too large (ensembles~EF-5, UF=5, and~SF-5 where $m_s/m_\ell=5$); the third result has the correct mass (SF-P). 
}
\end{figure}

This fit produces estimates for the $b$-quark's $\msb$ mass for each of the gluon ensembles:
\begin{equation}
    \mmsb_b^{(s)}\big(\mmsb_b^{(s)}\big) = 
    \rtag{\begin{cases}
        4.1192(62) & \mbox{SF-5 (0.059~fm)} \\
        4.1230(61) & \mbox{SF-P (0.057~fm)} \\
        4.1618(45) & \mbox{UF-5 (0.044~fm)} \\
        4.1833(28) & \mbox{EF-5 (0.033~fm)} \\    
    \end{cases}}
\end{equation}
These results are fit with \eq{eq:fit2} and extrapolated to zero lattice spacing (Fig.~\ref{fig:mbmb_a2}).
This is an excellent fit, \rtag{with a $\chi^2$ per degree of freedom of~0.8 for four degrees of freedom~($p=0.5$).} 
From it, we obtain our main result for the $\msb$ mass of the $b$-quark (without QED corrections),
\begin{equation}
    \label{eq:mbmb4}
    \mmsb_b(\mmsb_b,n_f=4) = \mbmbfour~\mathrm{GeV},
\end{equation}
which is equivalent to the $n_f=5$ result,
\begin{equation}
    \label{eq:mbmb5}
    \mmsb_b(\mmsb_b,n_f=5) = \mbmbfive~\mathrm{GeV}.
\end{equation}

\rtag{
\begin{table}
    \caption{Contributions to the ($1\sigma$) uncertainty in $\mmsb_b(\mmsb_b,n_f=5)$ (\eq{eq:mbmb5}) as a percentage of the mean value, in decreasing order of importance. These uncertainties are summed in quadrature to obtain the total uncertainty~\rtag{\cite{[{See the discussion of error budgets in Appendix~A of: }]Bouchard:2014ypa}}. QED corrections introduce a further uncertainty of 0.01\% that is not listed in this table. \rtag{Other contributions to the error budget are significantly smaller than~0.01\% and therefore negligible~($<0.003\sigma$).}}
    \label{tab:error-budget}
        \begin{ruledtabular}
            \rtag{\begin{tabular}{rl}
   $g_\alpha$, $g_{a^2,\alpha}$, $g_{c,a}$ (\eq{eq:xi_alpha}) &          0.104\% \\
                       $a\Lambda_n(m_h)\to 0$ (\eq{eq:aLamn}) &          0.068\% \\
               $\tilde g_m$, $\tilde g_{a^2m}$ (\eq{eq:xi_m}) &          0.059\% \\
                                 $\sigma_u$ (\eq{eq:sigma_u}) &          0.038\% \\
                                 $m_{\eta_b}$ (\eq{eq:metab}) &          0.023\% \\
                                      $r_n$ (\eq{eq:rn-tpth}) &          0.023\% \\
                $(m_c/m_h)^2$ correction (\eq{eq:nonzero-mc}) &          0.021\% \\
                 $a^{-1} = (w_0/a)/w_0$ (Table~\ref{tab:cfg}) &          0.019\% \\
                      $a\Lambda_{\eta_b}\to 0$ (\eq{eq:fit2}) &          0.017\% \\
                              $R_n$, $am_{\eta_h}$ statistics &          0.007\% \\[1.25ex]   
                                                        Total &          0.150\% \\
            \end{tabular}}
        \end{ruledtabular}
    \end{table}}

The error budget for $\mmsb_b(\mmsb_b, n_f=5)$ is listed in Table~\ref{tab:error-budget}. \rtag{The bulk~(92\%) of the uncertainty in the final result comes from the first three entries in the table. That is, it comes from: 
\begin{enumerate}
\item the priors for the parameters $g_\alpha$, $g_m$\,\ldots\,in the formulas used to correct $\almsb(\mu)$ and $\mmsb_h(\mu)$ for the de-tuned sea-quark masses in ensembles~SF-5, UF-5, and~EF-5; and 
\item the priors for the coefficients of the $(a\Lambda_n(m_h)/\pi)^{2j}$ corrections to $R_n^{(s)}(m_n)$.
\end{enumerate}
} 

\rtag{The error budget can be used to examine the sensitivity of the uncertainty in $\mmsb_b(\mmsb_b, n_f=5)$ to potential changes in the widths of the priors used by the fit. For example, we might assume that doubling the uncertainty in the lattice spacing~$a^{-1}$ would double its contribution to the error budget ($0.019\%\to0.038\%$), thereby increasing the total uncertainty by~$0.024\sigma$. This is slightly larger than the actual increase~($0.021\sigma$, determined by redoing the fit with the larger $a^{-1}$~uncertainty), but, in either case, the impact is negligible.} 

\rtag{Using the error budget in this way can significantly overestimate the impact of a change for some priors. For example, as noted above, doubling the width of the priors for the coefficients of the $(a\Lambda_n(m_h)/\pi)^{2j}$ corrections increases the final uncertainty by $0.14\sigma$ which is about half what the error budget suggests ($0.27\sigma$). The difference arises because the coefficients are constrained by the simulation data as well as the priors, thereby muting the impact of broadening the priors (especially for~$j=1$). In such cases the error budget gives an upper bound for the impact of a change. Doubling the widths of all the priors associated with the first three lines of the energy-budget table (Table~\ref{tab:error-budget}) increases the total uncertainty by~$0.8\sigma$ (and reduces the mass by~$0.25\sigma$). }

\begin{figure}
    \begin{center}
    \includegraphics[scale=0.9]{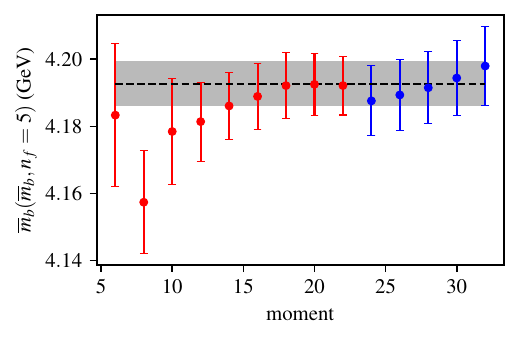}    
    \end{center}
\caption{\label{fig:mbmb_n} Results for the $b$-quark mass $\mmsb_b(\mmsb_b,n_f=5)$ obtained from each moment~$n$ separately. The dashed line and grey band correspond to the result obtained from the joint analysis of all moments from $n=6$ through $n=22$ together (red data points, \eq{eq:mbmb5}).}
\end{figure}

\begin{figure}
    \begin{center}
    \includegraphics[scale=0.9]{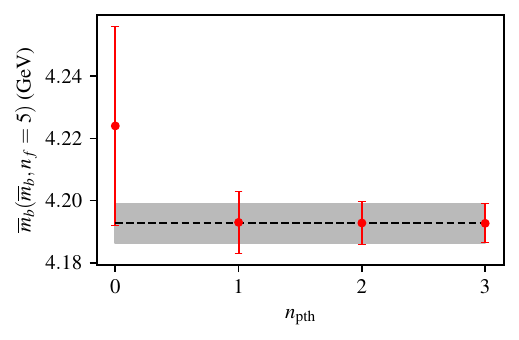}    
    \end{center}
\caption{\label{fig:convergence} Results for the $b$-quark mass $\mmsb_b(\mmsb_b,n_f=5)$ obtained from fits where the first $n_\mathrm{pth}$ coefficients in~$r_n$ (\eq{eq:rn-tpth}) come from perturbation theory and the remainder are replaced by fit parameters with priors $0\pm r_n^\mathrm{max}$ (\eq{eq:rnmax}). The dashed line and grey band correspond to the result obtained from the full analysis (\eq{eq:mbmb5}).} 
\end{figure}

We tested our analysis and results in a variety of ways, beyond the various checks on the fit priors detailed  above:
\begin{itemize}
    \item \textit{Individual moments:} Our final result is based on a joint analysis of all nine moments with~$6\le n\le 22$. It is possible to analyze the moments individually to obtain an $\mmsb_b(\mmsb_b,n_f=5)$ for each~$n$ separately. The results, which are shown in Fig.~\ref{fig:mbmb_n}, agree well with that from the joint analysis (\eq{eq:mbmb5}). The uncertainties are largest for smaller~$n$, where $am_h$~errors are large.\footnote{Nevertheless a joint fit to the moments with $n\le14$ gives \rtag{4.1916(74)~{GeV}} for $\mmsb_b(\mmsb_b,n_f=5)$, which is very close to the joint result from all moments with $n\le22$ (\eq{eq:mbmb5}).} Errors are also larger for $n>22$ because we do not have values for the third-order perturbation-theory coefficients $r_{n3}$ for those moments. The most accurate results come from the intermediate values~$14\le n\le 22$, where the moments are nonrelativistic and perturbation theory is known through third order. The masses from these intermediate~$n$s are in close agreement with the joint result and only slightly less accurate. 

    \item \textit{Larger masses:} We restrict the valence-quark mass~$m_h$ to be larger than~$0.65 m_b$ because higher masses are more perturbative (because $\mu_n$ is larger and $\almsb(\mu_n)$ smaller). Restricting $m_h$ to masses larger than $0.8m_b$ gives a $b$-quark mass of~\rtag{$4.1931(71)$}~GeV, in good agreement with \eq{eq:mbmb5}. 

    \item \btag{\textit{Smaller $am_h$:} Dropping the configuration sets with the largest lattice spacings (SF-5 and SF-P) reduces the maximum value for $am_h$ from~1.21 to~0.88. This gives a final result of $4.1896(71)$~GeV, which agrees well with \eq{eq:mbmb5}.}

    \item \textit{Scale independence:} We use the renormalization scales $\mu_n$ in Table~\ref{tab:rn-pth} for our perturbative expansions of the~$r_n$ (\eq{eq:rn-tpth}). Setting $\mu_n=\overline m_h(\mu_n)$ for all~$n$ instead gives a final result of~\rtag{$4.1918(64)$}~GeV for $\mmsb_b(\mmsb_b,n_f=5)$, which agrees well with our final result above.

    \item \textit{Higher-order perturbation theory:} We know the perturbative coefficients for the~$r_n$ through order~$\almsb^3(\mu_n)$, but also include a fourth-order term whose coefficient is a fit parameter with prior $0\pm r_n^\mathrm{max}$ (\eq{eq:rnmax}). This is to account for errors caused by truncating the perturbative expansion. To test this strategy we replace the third-order coefficient~$r_{n3}$ by a fit parameter with prior $0\pm r_n^\mathrm{max}$ and refit. This gives a $b$~mass of \rtag{$4.1928(69)$} which is very close to our final result (\eq{eq:mbmb5}) but with an uncertainty that is 10\%~larger. The close agreement is possible because the fit uses the lattice data to estimate the $r_{n3}$~coefficients. For example, it gives values
    \begin{equation}
        \rtag{r_{n3}\Big|_\mathrm{fit} = 
        \begin{cases}
            0.46(26) & \mbox{for $n=6$} \\
            0.157(97) & \mbox{for $n=14$}\\
            0.120(82) & \mbox{for $n=22$}, \\
        \end{cases}}
    \end{equation}
    which have large uncertainties but agree well with the corresponding values from Table~\ref{tab:rn-pth}: $0.463$, $0.171(48)$, and $0.120(10)$, respectively. The fact that the large uncertainties in these estimates increase the uncertainty on the $b$~mass by only~10\% confirms that the mass depends only weakly on third-order perturbation theory. Our lattice data and fits give no indication that fourth or higher orders are important at our level of precision.

    \item \textit{Convergent perturbation theory:} We examine the convergence of perturbation theory directly by refitting the lattice data with $r_n$ in \eq{eq:fit1} replaced by
    \begin{equation}
        r_n\to 1 + \sum_{j=1}^{N_\mathrm{r}} \tilde r_{nj} \,\almsb^j,     
    \end{equation}
    where $N_\mathrm{r} = 4$, 
    \begin{equation}
        \tilde r_{nj} \equiv 
        \begin{cases}
            r_{nj} & \mbox{for $j\le n_\mathrm{pth}$} \\
            0\pm r_n^\mathrm{max} & \mbox{for $j> n_\mathrm{pth}$} \\
        \end{cases}
    \end{equation}
    and $n_\mathrm{pth}$, the number of coefficients taken from perturbation theory,  ranges from~0 to~3. The results from these fits are shown in Fig.~\ref{fig:convergence}. Perturbation theory converges quickly: the uncertainties shrink with each order of perturbation
    theory, and results at each order agree well with our final result (dashed line and grey band). This is further evidence that our estimates of the truncation error in \eq{eq:rn-tpth} are reliable. 
    
    \item \textit{Evolution:} As an additional test 
    of perturbation theory we replace the leading coefficients $\beta_0$ and $\gamma_0$ in the coupling and mass evolution equations (Eqs.~(\ref{eq:evol-alpha}, \ref{eq:evol-m})) by fit parameters, each with a prior of~$0.5\pm1$, and refit the lattice data. This modified fit gives (nonperturbative) values for these parameters,
    \rtag{\begin{equation}
        \gamma_0 = 0.304(8) \quad\quad \beta_0 = 0.70(5),
    \end{equation}}
    that are consistent with the perturbative values,~0.318 and~0.663, respectively.
    Scale $\mu_n$ ranges from~2.7 to 7.8~GeV in the fit to \eq{eq:fit1}. The QCD
    coupling in the fit varies between~0.265 and~0.186 over this range.
     
\end{itemize} 

\section{QED Corrections}
\label{sec:qed}  
\rtag{In this section we calculate the QED correction to the results for~$\mmsb_b(\mmsb_b)$ from the previous section. The size of the QED correction depends upon the scheme used for setting QCD's parameters (quark masses and lattice spacing) in the absence of QED\,---\,different choices lead to slightly different results. Our QCD scheme is described in Sec.~\ref{sec:simulations}.\footnote{\rtag{This QCD scheme was chosen to match those used in our earlier papers~\cite{Chakraborty:2014aca,Hatton:2020qhk,Hatton:2021syc}. This allows us here to reuse results from those papers.}}}

\rtag{We calculate the QED correction to leading order in~$\alpha_\mathrm{QED}$ using the quenched approximation, where only QED interactions that involve the valence quarks are included.  Quenched results are believed to dominate other contributions from QED; QED corrections to $m_{\eta_b}$ that involve sea~quarks, for example, are suppressed by $\almsb^2(\Lambda_{\eta_b})$ where, again, $\Lambda_{\eta_b}\approx0.3 m_b$ is the typical valence-quark momentum in an $\eta_b$~meson.\footnote{\rtag{See Sec.~II.C in~\cite{Hatton:2020qhk} for more information about quenched QCD and heavy-quark masses.}} In our scheme the QED correction for the lattice $b$~mass is the shift in the quark~mass required to keep the $\eta_b$~mass the same, and equal to its experimental value, in both QCD and QCD+QED.\footnote{\rtag{Normally the sea-quark masses used in the 
QCD simulation are slightly different from those in the QCD+QED simulation. The $b$~mass in our scheme, however, depends only weakly on the sea-quark masses and therefore is largely unaffected by QED corrections to these masses. For example, from Table~\ref{tab:etasqed}, the QED correction to $m_{\eta_s}^2$ (at fixed valence-quark mass) is about~0.06\%. Assuming the corresponding $\delta m_{uds}^\mathrm{sea}/m_s$ is about~0.0006, such a correction would shift the $b$~mass in~\eq{eq:fit2} by less than~0.001\%, which is negligible~($<0.01\sigma$) compared to the uncertainty in our final result. A similar analysis shows that QED corrections to the mass of $c$~quarks in the sea are also negligible~($<0.02\sigma$).}}}

QED corrections combine self-energy and QED quark-antiquark interaction contributions which have opposite sign for electrically neutral mesons. For charmonium mesons, the self-energy term is larger and the combined effect is to raise the meson mass at fixed quark mass. For bottomonium each QED contribution is reduced by the smaller electric charge and there is a stronger cancellation as the interaction contribution is relatively larger~\cite{Hatton:2020lnm}. We therefore expect a much smaller QED correction to the $b$-quark mass in our scheme. 

The corrections for the $b$-quark mass were analyzed in \cite{Hatton:2021syc}. Eqs.~(28) and~(33) from that paper can be combined to obtain the shift $\delta \mmsb_b$ due to QED in the $b$-quark's $\msb$ mass at $\mu=3$~GeV:\footnote{In the notation of~\cite{Hatton:2021syc}, this ratio is given by $R(\mmsb_b/\mmsb_c,Q=1/3)$ times $1 + \big(R(\mmsb_c(3), Q_c=0\to 2/3)-1\big)/4$.}
\begin{equation}
    \label{eq:dmb/mb-qed}
    \frac{\delta \mmsb_b(3~\mathrm{GeV},n_f=4)}{\mmsb_b(3~\mathrm{GeV},n_f=4)}\Bigg|_\qed = 
    -0.00007(10),
\end{equation}
or about $-0.01(1)$\%. Combining this result with our pure QCD result from~\eq{eq:mbmb4} gives
\begin{equation}
    \label{eq:mb3}
    \mmsb_b(3~\mathrm{GeV}, n_f=4)\big|_{\mathrm{QED}} = \mbthreefourqed~\mathrm{GeV}
\end{equation}  
which implies that
\begin{equation}
    \mmsb_b(\mmsb_b)\big|_{\mathrm{QED}} = 
    \begin{cases}
        \mbmbfourqed~\mathrm{GeV} & \mbox{for $n_f=4$} \\
        \mbmbfiveqed~\mathrm{GeV} &  \mbox{for $n_f=5$}.
    \end{cases}
\end{equation}
We include the $\mathcal{O}(\alpha_\mathrm{QED})$ corrections to the perturbative evolution equations and matching equations used here, with $\alpha_\mathrm{QED}(3~\mathrm{GeV})=1/133$~\cite{Jegerlehner:2019lxt}.

\section{Charm Quark Mass}
\label{sec:charm}

The only other lattice QCD determination of the $b$-quark mass that includes QED effects in the simulations
combines nonperturbative results for the ratio of the~$b$ to~$c$ quark masses with an accurate calculation of the $c$~quark mass obtained using an intermediate RI-SMOM renormalization scheme~\cite{Hatton:2021syc}. It gives $4.202(21)$~GeV for the QED-corrected value of $\mmsb_b(\mmsb_b,n_f=5)$, which agrees well with our new result but is three times less accurate. It is less accurate because of the uncertainty in the $c$-quark mass, which suggests that the $c$~mass might be improved by deriving it from the $b$~mass instead of the other way around.  

We combine our new determination of the $b$-quark mass in~\eq{eq:mb3} with the QED-corrected ratio~\cite{Hatton:2021syc}  
\begin{equation}
    \frac{\mmsb_b(3~\mathrm{GeV}, n_f=4)}{\mmsb_c(3~\mathrm{GeV}, n_f=4)}\bigg|_\mathrm{QED}
    = 4.586(12)
\end{equation}
to obtain a new value for the QED-corrected $c$-quark mass:
\begin{equation}
    \mmsb_c(3~\mathrm{GeV}, n_f=4)\big|_\mathrm{QED} = 
        \mcthreefourqed~\mathrm{GeV} 
\end{equation}
It agrees well with the QED-corrected result 0.9841(51)~GeV obtained in~\cite{Hatton:2020qhk}, but is significantly more accurate.

Coming down from the $b$~mass to the $c$~mass, using the separately determined (nonperturbative) ratio of $b$~and $c$~masses in this way, makes sense. As discussed in Sec.~\ref{sec:a-uncertainty}, the accuracy of quark mass determinations is often limited by the accuracy of the lattice spacing~$a$, but this uncertainty is suppressed for heavy-quark masses. And this suppression is stronger for heavier quarks\,---\,2--3~times stronger for $b$~masses than for $c$~masses in our analysis.  Provided $am_b$ errors are under control, as they are here, it is more accurate to determine the $c$~quark mass from the $b$~quark mass than vice versa.

\section{Strange Quark Mass}
\label{sec:smass}
\rtag{As in the previous section we can derive a mass for the strange quark using our new value for the $b$-quark mass together with a (nonperturbative) determination of the ratio~$m_b/m_s$ from lattice QCD. As discussed in Sec.~\ref{sec:simulations}, we take the mass ratio \eq{eq:mb_ms} from Ref.~\cite{FermilabLattice:2018est} after correcting for slight differences in the QCD scheme used there from what we use. This ratio does not include QED corrections. To add these in we set
\begin{eqnarray}
\label{eq:QEDrat} 
\left. \frac{m_b}{m_s}\right|_{\mathrm{QED}} &=& 53.93(12)\times \\
&&\left(1 -  \left.\frac{\delta m_s}{m_s}\right|_{\mathrm{QED}} + \left.\frac{\delta m_b}{m_b}\right|_{\mathrm{QED}}\right) \, .\nonumber
\end{eqnarray}
where $\delta m_b/m_b$ is given in \eq{eq:dmb/mb-qed} (at a scale of 3 GeV) and is a negligibly small adjustment.  A more important, but still very small, effect comes from $\delta m_s/m_s$.}

\rtag{We tune the $s$~mass so that the mass of the (fictitious) $\eta_s$~meson equals its physical value (Sec.~\ref{sec:simulations}). To do this we need to know $m_{\eta_s}$ from simulations with both QCD and QED.} In fact we find that the physical value of the $\eta_s$ mass is the same in QCD+QED as it is in QCD as defined in~\cite{Dowdall:2013rya}. This fact is demonstrated in Appendix~\ref{appendix:ms} using quenched QED on two $n_f=2+1+1$ gluon field ensembles generated by the MILC collaboration~\cite{MILC:2012znn} with lattice spacing values of~0.12~fm and~0.15~fm, and physical light quarks. We can then define $\delta m_s/m_s$ as the relative change in~$m_s$ needed to keep the $\eta_s$~mass fixed when QED is included. This is given (using leading-order chiral perturbation theory) by 
\begin{eqnarray}
\label{eq:ms-etas}
\left.\frac{\delta m_s}{m_s}\right|_{\mathrm{QED}}&=& -\left.\frac{\delta m^2_{\eta_s}}{m^2_{\eta_s}}\right|_{\mathrm{fixed}\, am_s,\,\mathrm{QED}} \\&=& -2 \left[R^0_{\mathrm{QED}}(am_{\eta_s}) - 1 \right] \nonumber
\end{eqnarray}
where 
\begin{equation}
\label{eq:R0def}
R^0_{\mathrm{QED}}(am_{\eta_s}) = \left.\frac{am_{\eta_s}[{\mathrm{QCD+QED}}]}{am_{\eta_s}[{\mathrm{QCD}}]}\right|_{\mathrm{fixed}\,am_s}\, .
\end{equation}
This gives $\delta m_s/m_s$ values of $-0.000545(18)$ on ensemble VC-P and $-0.000579(5)$ on ensemble C-P (see Table~\ref{tab:etasqed} of Appendix~\ref{appendix:ms}). These correspond to fractional mass shifts under QED at different scales, because of the differing lattice spacing values. We find that the tuned lattice masses with the HISQ action are close in value to $\msb$~masses at a scale approximately~$1.5/a$ (see tuned $m_b$ values in Table~\ref{tab:cfg} or tuned $m_s$~values in~\cite{Chakraborty:2014aca}). This then seems an appropriate $\mu$ value. For VC-P this scale is~1.96~GeV and for C-P it is~2.44~GeV. 

Since $m_b/m_s$ is scale-invariant in both QCD and QCD+QED, the combination $\delta m_s/m_s - \delta m_b/m_b$ should be scale-invariant but we must combine $\delta m_s/m_s$ and $\delta m_b/m_b$ at the same scale. We therefore run $\delta m_b/m_b$ down from 3 GeV to the scales corresponding to the $\delta m_s/m_s$ calculation. Because masses run faster in QCD+QED than in QCD this increases $\delta m_b/m_b$. We find
\begin{equation}
    \label{eq:rundeltamb}
    \left.\frac{\delta m_b}{m_b}\right|_{\mathrm{QED}} = 
    \begin{cases}
        0.00002(10) & \mbox{for $\mu=  2.44$~GeV} \\
        0.00011(10) & \mbox{for $\mu=  1.96$~GeV} 
    \end{cases}
\end{equation}
and the combination 
\begin{equation}
    \label{eq:runcomb}
    \left(\left.\frac{\delta m_b}{m_b} -\frac{\delta m_s}{m_s}\right)\right|_{\mathrm{QED}} = 
    \begin{cases}
        0.00060(10) & \mbox{for $\mu=  2.44$~GeV}  \\
        0.00066(10) & \mbox{for $\mu=  1.96$~GeV} 
    \end{cases}
\end{equation}
The combination is indeed scale-invariant within the uncertainty quoted here but we take an additional conservative uncertainty of~0.0003 to allow for a factor of~2 variation in the assignment of~$\mu$ as~$1.5/a$. 

Taking the central value from the finer lattice in \eq{eq:runcomb}
and substituting into \eq{eq:QEDrat} gives 
\begin{eqnarray}
\left. \frac{m_b}{m_s}\right|_{\mathrm{QED}} &=& 53.94(12)\times (1+0.00060(10)(30)) \nonumber \\
&=& 53.97(12)(2),
\end{eqnarray}
where we have combined the QED correction uncertainties into the second error of the result. 
We conclude that the effect of QED on the ratio is very small and the corresponding impact on its uncertainty is negligible. 

We can now determine $m_s$ by combining this ratio with our $b$-quark mass values from~\eq{eq:mb3}. This gives 
\begin{align}
\label{eq:msval}
\left.\overline{m}_s(3\,\mathrm{GeV}, n_f=4)\right|_{\mathrm{QED}} &= \msthreefourqedmean{}(18)_{\mmsb_b}(19)_{m_b/m_s}~\mathrm{MeV} \nonumber \\ 
&= \msthreefourqed~\mathrm{MeV}.
\end{align}

\section{Conclusions}
\label{sec:conclusions}

\begin{figure}
    \begin{center}
    \includegraphics[scale=0.9]{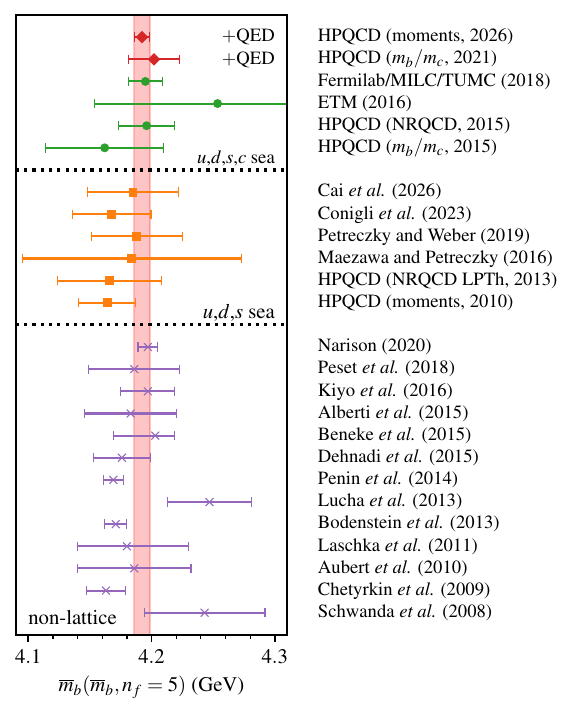}    
    \end{center}
\caption{\label{fig:mbcomp} Determinations of the $\msb$ $b$-quark mass from lattice QCD simulations with $n_f=4$ (green circles)~\cite{FermilabLattice:2018est,ETM:2016nbo,Colquhoun:2014ica,Chakraborty:2014aca} and $n_f=3$ (orange squares)~\cite{Cai:2026xja,Conigli:2023rod,Petreczky:2019ozv, Maezawa:2016vgv,Lee:2013mla,McNeile:2010ji} quark flavors in the sea, and from non-lattice analyses (purple crosses)~\cite{Narison:2019tym,Peset:2018ria,Kiyo:2015ufa,Alberti:2014yda,Beneke:2014pta,Dehnadi:2015fra,Penin:2014zaa,Lucha:2013gta,Bodenstein:2011fv,Laschka:2011zr,BaBar:2009zpz,Chetyrkin:2009fv,Belle:2008fsc}. The top two results are from lattice simulations that include QCD and (quenched) QED, with $n_f=4$ flavors (red diamonds)\,---\,this paper and~\cite{Hatton:2021syc}. Results from the HPQCD collaboration use multiple methods to determine the quark mass: moments of heavy-quark current-current correlators, nonperturbative determinations of $m_b/m_c$ together with earlier determinations of $\mmsb_c$, moments of current correlators calculated using NRQCD rather than HISQ for heavy quarks, and QCD lattice perturbation theory to relate the bare lattice NRQCD quark mass to its $\msb$ mass. The red band corresponds to the results from this paper (\eq{eq:mbmbfinal}).}
\end{figure}

Our final result for the $\msb$ $b$-quark mass is the most accurate to date and one of only two lattice results that includes (quenched) QED corrections in the simulations:
\begin{equation}
    \label{eq:mbmbfinal}
    \mmsb_b(\mu, n_f=5)\big|_\mathrm{QED} = 
    \begin{cases}
            \mbmbfiveqed~\mathrm{GeV}  & \mu = \mmsb_b(\mu) \\
            \mbtenfiveqed~\mathrm{GeV}  & \mu = 10~\mathrm{GeV}
    \end{cases}
\end{equation}
The $\mu=\mmsb_b(\mu)$ result compares well with the world average from the Particle Data Group (PDG), $4.183(7)$~GeV~\cite{ParticleDataGroup:2024cfk}. In Fig.~\ref{fig:mbcomp} we compare our result (red band) with recent results from lattice QCD. We also compare it with the non-lattice results used by the PDG in their average. Results are generally  consistent across a wide variety of techniques for determining the mass, which gives us confidence in the final results. In particular, the most accurate previous lattice result, from the Fermilab/MILC/TUMQCD collaborations~\cite{FermilabLattice:2018est}, agrees very well with our result, and is based on fitting heavy-quark effective field theory (HQET) to lattice results for light-heavy pseudoscalar-meson masses\,---\,a completely different approach from that taken here, with very different systematic errors. This analysis, in common with earlier calculations, did not include QED corrections but estimated their effect based on phenomenological arguments~\cite{Goity:2007fu,Davies:2010ip}.

Earlier results from the HPQCD collaboration also relied upon a variety of techniques. These included using the nonrelativistic QCD (NRQCD) action for heavy quarks, rather than the HISQ action, when calculating moments of the current-current correlators~\cite{Colquhoun:2014ica}. This approach avoids large $am_b$ errors entirely, and works well for higher moments ($n\ge 14$), where the heavy quarks are nonrelativistic. Another approach also used the NRQCD action but related the tuned lattice bare quark mass~$m_b$ directly to the $\msb$ mass~$\mmsb_b(\mmsb_b)$ using lattice perturbation theory~\cite{Lee:2013mla}. 

Our result for the $b$~mass, when combined with nonperturbative lattice determinations of the ratio of the $b$~and $c$~masses, implies a new value for the QED-corrected $c$~mass:
\begin{equation}
    \mmsb_c(\mu, n_f=4)\big|_\mathrm{QED} = 
    \begin{cases}
        \mcmcfourqed~\mathrm{GeV} & \mu=\mmsb_c(\mu) \\
        \mcthreefourqed~\mathrm{GeV} & \mu=3~\mathrm{GeV}.
    \end{cases}
\end{equation}
This estimate is the most  accurate to date from lattice QCD. It agrees well with the $\mu=3$~GeV QED-corrected result 0.9841(51)~GeV obtained in~\cite{Hatton:2020qhk}, and with the PDG average~1.2730(46)~GeV for $\mu=\mmsb_c(\mmsb_c)$~\cite{ParticleDataGroup:2024cfk}. 

We also combined our new $b$~mass with recent lattice results for 
$m_b/m_s$~\cite{FermilabLattice:2018est} to obtain a new value for the QED-corrected $s$-quark mass:
\begin{equation}
   \mmsb_s(\mu, n_f=4)\big|_\mathrm{QED} = 
    \begin{cases}
        \mstwofourqed~\mathrm{MeV} & \mu=2~\mathrm{GeV} \\
        \msthreefourqed~\mathrm{MeV} & \mu=3~\mathrm{GeV}.
    \end{cases}    
\end{equation}
This is the most accurate determination by any method. The $\mu=2$~GeV result agrees well with the current average~93.5(8)~MeV from the PDG~\cite{ParticleDataGroup:2024cfk}. It also agrees well with the most accurate previous lattice result~92.47(69)~MeV~\cite{FermilabLattice:2018est}. 

\rtag{We used quenched QED simulations to correct these masses for QED effects. The corrections using our QCD scheme (Sec.~\ref{sec:simulations}) are small, at the scales used above, when compared with the total uncertainties: for example,
\begin{equation} 
    \frac{\delta \mmsb_q(3\,\mathrm{GeV}, n_f=4)}{\sigma_{\mmsb_q(3\,\mathrm{GeV},n_f=4)}}\bigg|_\mathrm{QED} = 
    \begin{cases}
    -0.03(5) & \mbox{$q=b$} \\
    -0.51(5)  & \mbox{$q=c$} \\
    -0.22(11) & \mbox{$q=s$}.
    \end{cases}
\end{equation} 
The impact of the QED corrections on our results for the quark masses ranges from small for the $c$~quark to negligible for the $b$~quark.}

Ref.~\cite{Lepage:2014fla}, written eleven years ago, analyzed the extent to which lattice QCD could achieve precisions adequate to meet the needs of the proposed International Linear Collider (ILC) for high-precision measurements of Higgs couplings. The critical parameters from QCD are the masses of the $b$~and $c$~quarks, and the strong coupling constant, which at that time were known to within 0.69\%, 0.61\%, and~0.59\%, respectively, from lattices with lattice spacings as small as 0.044~fm~\cite{Chakraborty:2014aca,McNeile:2010ji}.\footnote{Specifically, the uncertainties are for the following: $\mmsb_b(10~\mathrm{GeV}, n_f=5)$, $\mmsb_c(3~\mathrm{GeV}, n_f=4)$, and $\almsb(M_Z,n_f=5)$, respectively.}  Ref.~\cite{Lepage:2014fla} used simple simulations to estimate that adding new results from lattices with lattice spacing~0.032~fm would reduce the uncertainties on the $b$~and $c$~quark masses to {0.30\% and 0.53\%,} respectively. In fact, we have shown here that those uncertainties are reduced to~0.26\% and~0.35\%, respectively, which is accurate enough to meet the proposed ILC target precisions for Higgs decays to $b\bar b$ and $c\bar c$.\footnote{See Table~I in~\cite{Lepage:2014fla}. The Higgs coupling to gluons has not been significantly improved in the last decade, but was already close to being accurate enough back then.}  These also exceed the projected precisions for the quark masses used to analyze the impact of the Future Circular Collider (FCC) and other future colliders in~\cite{Freitas:2019bre, deBlas:2019rxi}.\footnote{It is important to improve the accuracy of $\almsb(M_Z,n_f=5)$ (currently about 0.6\%), as well as the quark masses, since the coupling is needed to evolve the quark masses to the Higgs mass. Evolving our result for the $b$-quark mass from $\mu=10$~GeV to the Higgs mass increases its uncertainty from 0.26\% to 0.5\%. Reducing the coupling's uncertainty by~40\% would reduce the $b$~mass's uncertainty at the Higgs scale to 0.3\%.} 

Our new results are more accurate than might have been expected a decade ago due to three features of our analysis. One is the strong suppression of $a m_b$ errors (by $(v/c)^4$) for the higher (non-relativistic) moments (Sec.~\ref{sec:finite-a}); the earlier analyses used only moments with $n=4$--10. The second feature, evident from Table~\ref{tab:rn-pth}, is that the higher moments are also much less dependent on the value (and accuracy) of the strong coupling constant, because their perturbative coefficients are substantially smaller (Sec.~\ref{sec:moments}). Therefore errors from truncating perturbation theory are smaller. The third feature is the insensistivity of our result for the $b$~mass to the lattice spacing (Sec.~\ref{sec:a-uncertainty}) which implies that very accurate $c$-quark mass estimates are obtained by combining $b$-quark mass estimates with nonperturbative results for the ratio of the $b$~to $c$~masses (Sec.~\ref{sec:charm}). 

The error budget (Table~\ref{tab:error-budget}) for the $b$~mass suggests that the total error could be reduced by almost half with simulations using realistic sea-quark masses for all lattice spacings\,---\, that is for 0.032~fm and~0.044~fm, as well as for~0.057~fm. Such simulations are feasible with today's computing resources. Reducing the smallest lattice spacing to~0.023~fm would provide a further significant improvement. 

\section*{Acknowledgements}
We thank J.~Koponen and A.~Lytle for useful input at early stages of this work and we are grateful to the MILC collaboration for making publicly available their gauge configurations and code.
This work used the DiRAC Data Intensive Service (CSD3) at the University of Cambridge, managed by the University of Cambridge Information Services on behalf of the Science and Technology Facilities Council (STFC) DiRAC HPC Facility (www.dirac.ac.uk). The DiRAC component of CSD3 at Cambridge was funded by BEIS, UKRI and  STFC capital funding and STFC operations grants. DiRAC is part of the UKRI Digital Research Infrastructure.
We are grateful to the CSD3 support staff for assistance.
Funding for this work came from STFC grant ST/T000945/1.

\begin{appendix}

\section{Nonrelativistic effective field theory for HISQ}
\label{app:nrHISQ}
The nonrelativistic structure of the HISQ action was analyzed in~\cite{Follana:2006rc}. Here we repeat that analysis but keep terms to all orders in~$am$ so we can examine the behavior when the quark mass is very large~($am>1$)~\cite{Monahan:2012dq}. To simplify the formulas, we work in lattice units, where $a=1$. Following our earlier analysis, the nonrelativistic tree-level reduction of the HISQ action is trivial since the action is identical in structure to Dirac theory: for example, the equations of motion are 
\begin{equation}
	\left(\tilde\Delta\cdot\gamma +  m_0 \right)\,\Psi = 0
\end{equation}
where $\tilde\Delta$ is the improved derivative with fattened links. The nonrelativistic expansion is the same as in the continuum but with continuum derivatives $D$ replaced by $\tilde\Delta$s:
\begin{align}
	\label{nrexp1}
	\Big(
	\tilde\Delta_t + m_0 &- \frac{\tilde\Deltav^2}{2m_0}
	 -\frac{(\tilde\Deltav^2)^2}{8m_0^3}
	 -\frac{g\sigma\cdot\Bv}{2m_0} \nonumber \\
     &+ \frac{ig(\tilde\Deltav\cdot\Ev - \Ev\cdot\tilde\Deltav)}{8m_0^2} + \cdots
	\Big)\,\psi = 0
\end{align}
Ignoring all errors except $am$~errors (and working to lowest order in $\alpha_s$), we can replace 
\begin{align}
    \label{eq:improved-Delta-v}
    \tilde\Deltav &\to \sinh(\Dv) - (1+\epsilon) \frac{\sinh^3(\Dv)}{6} \nonumber \\
    &\to \Dv + \mathcal{O}(a^4\Dv^5) \\
	\tilde\Delta_t &\to \sinh(D_t-m) - (1+\epsilon) \frac{\sinh^3(D_t-m)}{6}
    \label{eq:improved-Delta-t}
\end{align}
where again $D$ is the continuum covariant derivative, and where we have made replacement $\psi \to \exp(-mt)\psi$ with $m$ equal to the exact (pole) mass of the free quark. With these definitions $D_t$ is of order the binding energy and therefore much smaller than~$m$. So we can expand $\tilde\Delta_t$ in powers of $D_t$:
\begin{equation}
    \label{eq:Deltat}
	\tilde\Delta_t = W(m) + X(m) D_t + Y(m) D_t^2 + \cdots
\end{equation}
where
\begin{align}
    \label{Wdefn}
	W(m) &\equiv -\sinh(m) +\frac{1+\epsilon}{6}\,\sinh^3(m) %
     \\
	X(m) &\equiv \cosh(m)\Big(1-\frac{1+\epsilon}{2}\,\sinh^2(m)\Big) 
        \\
	Y(m) &\equiv \frac{\epsilon}{2} \,\sinh(m) + \frac{3(1+\epsilon)}{4}\,\sinh^3(m)  
\end{align}
Substituting back into \eq{nrexp1} gives
\begin{align}
    \label{nrexp3}
	&X \Big(
	D_t + \frac{Y}{X}\,D_t^2 + \frac{W+m_0}{X} - \frac{\Dv^2}{2m_0X}
	 -\frac{\Dvfour}{8m_0^3X}  \\ \nonumber
	 & -\frac{g\sigma\cdot\Bv}{2m_0X}
	 + \frac{ig(\Dv\cdot\Ev - \Ev\cdot\Dv)}{8m_0^2X} + \cdots
	\Big)\,\psi = 0.
\end{align}

\begin{figure}
    \begin{center}
    \includegraphics[scale=0.9]{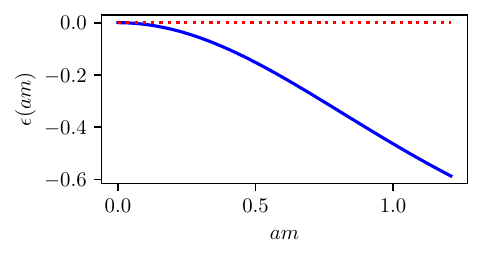}
    \includegraphics[scale=0.9]{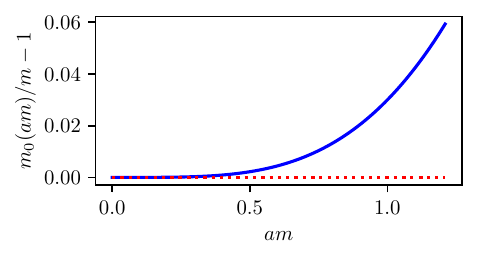}       
    \end{center}
\caption{\label{fig:eps-m0}Plots of $\epsilon(am)$ (top) and $m_0(am)/m - 1$ (bottom) as a function of quark mass~$am$.}
\end{figure}

We put this equation into a more canonical form by first choosing $m_0$ and $\epsilon$ so that:
\begin{align}
    m_0 &= - W(m) \\
    m &= m_0 X(m) \\
      &= - W(m) X(m)
\end{align}
where, again, $m$ is the quark mass (i.e., the pole mass). The first of these equations sets the rest energy to zero in the nonrelativistic expansion, and the second sets the kinetic mass equal to the quark mass.
The last equation can be rewritten as an equation for 
$\xi\equiv \sinh^2(m)(1+\epsilon)$,
\begin{equation}
    0 = \xi^2 - 8 \xi + 12 \Big(1 - \frac{2m}{\sinh(2m)}\Big),
\end{equation}
which has two solutions:
\begin{equation}
    \xi = 4 \pm \sqrt{4 + \frac{24m}{\sinh(2m)}}.
\end{equation}
Choosing the minus sign, so that $\epsilon$ is small when $m$ is small, we obtain the standard result for~$\epsilon$:
\begin{align}
    \label{eq:epsilon}
    \epsilon(m) &= -1 + \frac{4 - \sqrt{4 + 24m/\sinh(2m)}}{\sinh^2(m)} \\ 
    &= - \frac{27}{40}\,m^2 + \frac{327}{1120}\,m^4 - \cdots.
\end{align}
This equation and
\begin{align}
    m_0 &= -W(m) \label{m0eqn} \\ \nonumber
     &= \sinh(m) - (1+\epsilon(m))\sinh^3(m)/6
\end{align}
determine all the parameters in the nonrelativistic action. Fig.~\ref{fig:eps-m0} shows how $\epsilon$ and $m_0$ vary with quark mass.

The final simplification of \eq{nrexp3} is to use the field equation for $\psi$ to replace
\begin{equation}
	D_t \to \frac{\Dv^2}{2m_0X} = \frac{\Dv^2}{2m}
\end{equation}
in the $D_t^2$ term. The result is
\begin{align}
	\label{nrexp2}
	&X \Big(
	D_t - \frac{\Dv^2}{2m}
	 -X^2\,\frac{\Dvfour}{8m^3}
	+ \frac{mY}{X}\,\frac{\Dvfour}{4m^3}  \\ \nonumber
	 &-\frac{g\sigma\cdot\Bv}{2m} 
	+ X\,\frac{ig(\Dv\cdot\Ev - \Ev\cdot\Dv)}{8m^2} + \cdots
	\Big)\,\psi = 0.,
\end{align}

We can now write down our final result for the nonrelativistic 
effective field  theory corresponding to the HISQ action,
after undoing the $\psi\to\exp(-mt)\psi$ transformation:\footnote{We display every term in $\mathcal{L}_\mathrm{eff}^\mathrm{HISQ}$ through order~$1/m^2$. We also keep the $-\psi^\dagger (\Dv^2)^2\psi/8m^3$ term since it is competitive with the $1/m^2$ terms for mesons composed of a heavy quark and heavy anti-quark~\cite{Lepage:1992tx}.}
\begin{align} 
    \label{eq:nr-HISQ}
    \mathcal{L}_\mathrm{eff}^\mathrm{HISQ} &= X(m)\, \psi^\dagger \Big(
    D_t + m - \frac{\Dv^2}{2m}
    -c_K\,\frac{\Dvfour}{8m^3} \nonumber \\ 
    &-\frac{g\sigma\cdot\Bv}{2m} 
    + c_E\,\frac{ig(\Dv\cdot\Ev - \Ev\cdot\Dv)}{8m^2} \\ \nonumber
    &- c_E \,\frac{g\sigmav\cdot(\Dv\times\Ev - \Ev\times\Dv)}{8m^2}
    + \cdots
    \Big)\,\psi 
\end{align}
where
\begin{align}
    c_K &\equiv X^2(m) - 2mY(m)/X(m) \\ \nonumber
        &= 1 - \frac{9}{10}\,m^4 + \frac{29}{280}\,m^6 - \frac{4213}{67200}\,m^8 + \cdots \\ 
    c_E &= X(m) \\ \nonumber 
        &= 1 - \frac{3}{80}\,m^4 + \frac{23}{2240}\,m^6 - \frac{1241}{537600}\,m^8 + \cdots.
\end{align}

\begin{figure}
    \begin{center}
        \includegraphics[scale=0.9]{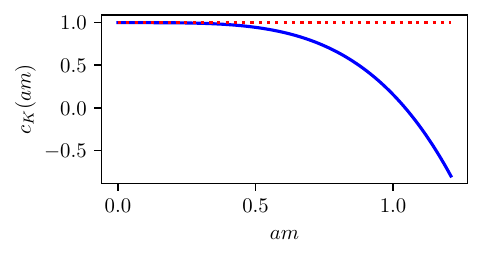}       
        \includegraphics[scale=0.9]{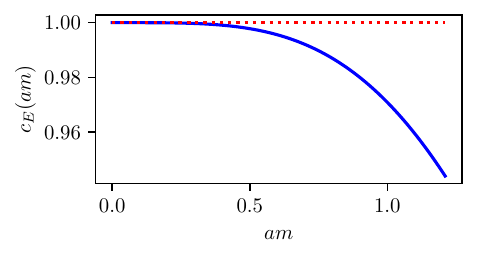}
    \end{center}
\caption{\label{fig:c_Xplots}Plots of $c_K(am)$ (top) and $c_E(am)$ (bottom) as a function of quark mass~$am$.}
\end{figure}

The effective Lagrangian is correct insofar as $c_K$ and $c_E$ are close to one; errors are proportional to $c_K-1$ and $c_E-1$. Fig.~\ref{fig:c_Xplots} shows how couplings $c_K$ and $c_E$ deviate from one over the range of quark masses $am$ we are using in this paper. Only $c_K$ shows large deviations but even these are only of~$\mathcal{O}(1)$ over this range. The deviations can become much larger at larger masses with 
\begin{align}
    m_0(m) &\to 2\sinh(m)/3 \to \infty \\
    \epsilon(m) &\to -1 + 2/\sinh(m)^2 \to -1\\
    c_E(m) &\to 3m/(2\sinh(m)) \to 0 \\
    c_K(m) &\to -2\sinh^2(m)/3 \to -\infty
\end{align}
as $m\to\infty$. $c_K$ diverges exponentially 
quickly in this limit; it is already equal to
$-1.6\times10^{8}$ by~$m=10$. 

The overall factor of $X(m)$ in \eq{eq:nr-HISQ} is a tree-level wave function renormalization: 
\begin{equation}
    \Psi_\mathrm{contin} \to \sqrt{X(m)}\, \Psi_\mathrm{HISQ}.
\end{equation}
Therefore, for example, the local HISQ current 
\begin{equation}
    J_\mu^\mathrm{loc}(x) \equiv X(m) \,\overline\Psi(x) \gamma_\mu \Psi(x)
\end{equation} 
has no tree-level $(am)^{2n}$ errors to all orders in~$n$.  Such factors cancel in our analysis, when we take the ratio $G_n/G_n^{(0)}$ in \eq{eq:Rn}.

\section{Propagator Poles, Ghost States}
\label{app:ghost}

\begin{figure}
    \begin{center}
        \includegraphics[scale=0.9]{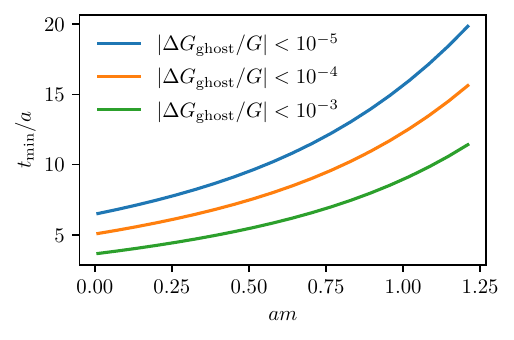}       
    \end{center}
\caption{\label{fig:tmin_m}$t_\mathrm{min}$ as a function of quark mass~$am$, where $t_\mathrm{min}$ is defined so that the fractional contribution from ghost states to the quark correlator $|\Delta G_\mathrm{ghost}/G|$ is smaller than $10^{-3}$ (bottom, green), $10^{-4}$ (middle, red), or $10^{-5}$ (top, blue) for $t\ge t_\mathrm{min}$.}
\end{figure}

The free-quark HISQ propagator with zero three-momentum and projected onto the $\gamma_t\to1$ spinor subspace is (in lattice units)
\begin{align}
    G^{-1} &= W(D_t) + m_0 \nl
        &= W(iE) + m_0
\end{align}
in Euclidean space, where $W$ is defined in \eq{Wdefn}. Defining 
\begin{equation}
    z \equiv \e^{iE},
\end{equation}
the free correlator in Euclidean $t$-space is 
\begin{align}
    G(t) &= \int\limits_{-\pi}^{\pi} \frac{dE}{2\pi}\, G(z(E))\, z(E)^t \nl 
    &= \oint\limits_{|z|=1} \frac{dz}{2\pi i z}\,G(z)\, z^t \quad\quad(t>0)
\end{align}
where the contour integral is counter clockwise around the origin, $t$ is a positive integer, and 
\begin{align}
    G^{-1}(z) &\equiv m_0(m) + (z - 1/z)/2 \nl
    &\quad\quad-(1+\epsilon(m))(z-1/z)^3/48 
    \label{Ginv}
\end{align}
Assuming $G^{-1}(z)$ has zeros at $z=z_n$, the integral gives
\begin{align}
    G(t) = \sum\limits_{|z_n|<1} A_n z_n^t
\end{align}
where $A_n$ is the residue of the pole in $G(z)/z$ at $z=z_n$. Finding the roots numerically, 
we obtain, for example, 
\begin{align}
    \label{exampleG}
    G(t) &= 
    \begin{cases} 
        1.00\,\e^{-0.1t}-0.20\,\e^{-1.61t}+0.17\,(-1)^t\,\e^{-1.65t} \\
        1.00\,\e^{-0.6t}-0.31\,\e^{-1.60t}+0.12\,(-1)^t\,\e^{-1.82t} \\
        1.06\,\e^{-1.2t}-0.53\,\e^{-1.75t}+0.08\,(-1)^t\,\e^{-2.17t} \\
        1.33\,\e^{-2.0t}-1.02\,\e^{-2.20t}+0.04\,(-1)^t\,\e^{-2.79t} \\
    \end{cases}
\end{align}
for $m=0.1$, 0.6, 1.2, and~2.0, respectively. The second and third states are algorithmic ghost states. The second violates unitarity because it has a negative amplitude. The third state is a doubler. The coefficient of the leading term equals $1/X(m)$, as expected from the tree-level wave function renormalization in~\eq{eq:nr-HISQ}.

The expressions  for the tree-level $G(t)$ with different $m$ values give us an idea as to how the ghosts might affect our simulations. The leading ghost violates unitarity. Our standard fitting code would have a hard time dealing with this because of the negative amplitude\,---\,we would be unable to get good fits. We can suppress the fractional contribution from ghost states $|\Delta G_\mathrm{ghost}(t)/G(t)|$ to be less than some fixed constant by restricting correlator fits to $t\ge t_\mathrm{min}$ for a $t_\mathrm{min}$ that depends on the quark mass (Fig.~\ref{fig:tmin_m}). The mass of the first ghost state gets closer and closer to the physical mass as $m$ increases, and therefore $t_\mathrm{min}$ increases with increasing mass. The fits for the $\eta_h$~masses   in this paper use $t$~values larger than~50, well beyond the $t_\mathrm{min}$~values in this figure. 

The zeros  of $G^{-1}(z)$ (\eq{Ginv}) come in pairs, $z_n$ and $-1/z_n$. Poles where $z_n<0$ correspond to doublers. The doublers for the states shown in~\eq{exampleG} are obtained by redoing that analysis but with $\gamma_t\to-1$  (instead of 1);  then $z_n\to-z_n$, introducing an extra factor of $(-1)^t$, and $A_n\to A_n$.

The anti-particle states  corresponding to the states in~\eq{exampleG} are obtained by taking $t<0$, $\gamma_t\to-1$, and replacing the integration variable by $z \to \tilde z \equiv 1/z$. There is one anti-particle state for each of the states that contribute when~$t>0$.

\section{QED and $m_u \ne m_d$ effects on $m_{\eta_s}$}
\label{appendix:ms}

In~\cite{Dowdall:2013rya} HPQCD performed 
a chiral-continuum extrapolation of lattice 
QCD results for $\pi$, $K$ and $\eta_s$~meson masses calculated on 
$n_f=2+1+1$~gluon field configurations including HISQ sea quarks and generated 
by the MILC collaboration~\cite{MILC:2012znn}. 
The ratio of $m_{\eta_s}^2$ to $(2m_K^2-m_\pi^2)$ in the continuum limit 
at physical quark masses and infinite volume was 
found to be: 
\begin{equation}
\label{eq:etas-mass2}
\frac{(m_{\eta_s})^2}{2(m_K)^2-(m_{\pi})^2} = 1.0063(64).
\end{equation}
This ratio can be calculated very robustly because it is insensitive to changes in light or strange 
quark masses, being 
exactly~1 in leading-order chiral perturbation theory. Indeed all of the results for the ratio from~\cite{Dowdall:2013rya} lie 
within the $\pm1\sigma$ band, including those with light quark masses as large as $m_s/5$ as well as those with significantly mistuned $s$ quark masses. 
To determine a physical value for~$m_{\eta_s}$ then all that is needed is to fix suitable values
from experiment for $m_{\pi}$ and~$m_K$. A combination of $\pi$ and $K$~masses 
that make the denominator insensitive to changes in 
$u$ and $d$~quark masses separately is needed, since the lattice calculation used~$m_u=m_d$. 
The combination should also allow for QED effects that are missing from the lattice calculation of~\cite{Dowdall:2013rya}.

\rtag{In a world without QED, all $\pi$~mesons would have the same mass (up to small corrections from the $(u-d)$~mass difference, which we allow for by increasing the uncertainty on $m_{\pi}$ to 0.32~MeV~\cite{Dowdall:2013rya}). We take the physical mass of the pion in that world to be~134.98(32)~MeV, which is the experimental mass of the neutral~$\pi^0$.} 
The physical value of~$m_K$ was taken as the root-mean-square average of 
the experimental~$K^+$ and $K^0$~meson masses (the correct value for 
the $m_u=m_d$~calculation) with an additional correction for QED effects. 
Taking
\begin{eqnarray}
(m_K^{\mathrm{phys}})^2 &=& \frac{1}{2}\left[ (m_{K^+}^2 + m_{K^0}^2) \right. \\
&-& \left. (1+\Delta_E)(m_{\pi^+}^2-m_{\pi^0}^2) \right], \nonumber 
\end{eqnarray}
gave a value for $m_K^{\mathrm{phys}}$ of 494.6(3) MeV, allowing the 
coefficient $\Delta_E$ that accounts for violations of Dashen's 
theorem~\cite{Dashen:1969eg} 
to be 0.65(50) (this then dominates the uncertainty in $m_K^{\mathrm{phys}}$)~\rtag{\cite{Dowdall:2013rya}}.

These values, combined with the ratio from Eq.~\eqref{eq:etas-mass2} 
yielded a physical value for $m_{\eta_s}$ of 
688.5(2.2)~MeV~\cite{Dowdall:2013rya}. The uncertainty of~0.3\% is dominated 
by those from statistics and extrapolation to~$a = 0$.  

Here we give results for the masses of the $\eta_s$, $K^+$, $K^0$, 
$\pi^+$ and~$\pi^0$ that include quenched QED effects as well as 
differing $u$ and $d$~quark masses. This allows us to check whether 
the value of the $\eta_s$~mass agrees with the result from 
the earlier pure QCD calculation or whether we need to change the physical 
value of the $\eta_s$ mass that we tune to in this calculation.  
Note that we do that by a comparison of QCD+QED to pure QCD on a 
lattice spacing by lattice spacing basis. This is because the 
ratio of Eq.~\eqref{eq:etas-mass2} is obtained in the continuum 
limit at infinite volume. Away from this point there are small 
discretisation effects (up to~0.6\% on very coarse lattices) and 
small QCD finite-volume effects (at most~0.2\%). These effects will 
cancel in a comparison of QCD+QED to QCD and so can be ignored.  

\begin{table}
\caption[hjk]{Parameters for the two $n_f=2+1+1$ HISQ ensembles 
used here, denoted VC-P and C-P. We give the 
lattice spacing in terms of $w_0$~\cite{Borsanyi:2012zs} in column 2. 
VC-P is ``very coarse'' ($a \approx$ 0.15 fm) and 
C-P, ``coarse'' ($a \approx$ 0.12 fm). 
Columns 3 and 
4 give the spatial and temporal extents of the lattices in 
lattice units. Columns 5 through 7 give the sea quark masses 
of the light, strange and charm quarks in lattice units. The 
light quark masses are close to their 
physical value; note that $m_u=m_d$. The final column gives the valence strange 
quark mass. This is more closely tuned than the sea mass to give the physical $\eta_s$ meson mass 
from~\cite{Dowdall:2013rya}. }
\begin{ruledtabular}
\begin{tabular}{llllllll}
Ensemble & $w_0/a$ & $L/a$ & $T/a$ & $am_l^{\mathrm{sea}}$ & $am_s^{\mathrm{sea}}$ & $am_c^{\mathrm{sea}}$ & $am_s^{\mathrm{val}}$  \\
\hline
VC-P & 1.1367(5) & 32 & 48 & 0.00235 & 0.0647 & 0.831 & 0.0677  \\
C-P & 1.4149(6) & 48 & 64 & 0.00184 & 0.0507 & 0.628 & 0.0527  
\end{tabular}
\end{ruledtabular}
\label{tab:ensembles}
\end{table}

We perform lattice QCD + quenched QED 
calculations (using the stochastic approach~\cite{Duncan:1996xy} and the $\mathrm{QED}_L$ scheme~\cite{Hayakawa:2008an}) with $m_u^{\mathrm{val}}\ne m_d^{\mathrm{val}}$ on two gluon field ensembles 
with parameters given in Table~\ref{tab:ensembles}. 
These two ensembles have 
$m_u=m_d=m_l$ in the sea with the light quark mass $m_l$ close to its physical value.  
The ensembles, labelled VC-P and~C-P, have two different 
values of the lattice spacing, around 0.15~fm and 0.12~fm respectively, 
so that we can test for any dependence on the lattice spacing in our 
findings.  Note that the lattice spacing is determined from a gluonic quantity, $w_0/a$, and the value of $w_0$ is fixed from the pure QCD quantity, $f_{\pi}$. We therefore expect the lattice spacing values to remain the same in QCD+QED~\cite{Hatton:2020qhk}.

In what follows quantities that  include the effect of strong isospin 
breaking will be denoted with a superscript~$u \neq d$. 
The valence masses that we use for $u$ and~$d$ in that case are given 
in Table~\ref{tab:Kqed}. They are chosen to have an average equal to 
the value of $am_l^{\mathrm{sea}}$ in Table~\ref{tab:ensembles}
and to have a $m_u/m_d$~ratio of 0.46~\cite{Basak:2016jnn}. 
When taking $am_u=am_d=am_l$ we use $am_l^{\mathrm{val}}=am_l^{\mathrm{sea}}$. 
The meson will be indicated with its correct charge assignment so that 
it is clear, even in the pure QCD calculation, which light quark and/or 
antiquark was included. Results for the case where quenched QED is included 
will be given in terms of the ratio to pure QCD, $R^0$, defined by
\begin{equation}
\label{eq:r0def}
R^0_{\mathrm{QED}}(X) \equiv \left. \frac{X[\mathrm{QCD+QED}]}{X[\mathrm{QCD}]}\right|_{\mathrm{fixed}\ am^{\mathrm{val}}}\, .
\end{equation} 
When quenched QED is included we use the correct electric
charges for the quark/antiquark pair even in the case where $m_u=m_d$.   
To obtain good precision for these ratios we average over opposite electric 
assignments to remove noise linear in~$e$~\cite{Blum:2007cy}. 

\begin{table}
\caption{Results for QCD $K$ meson masses in lattice units and 
QCD+QED/QCD ratios $R^0$ as determined on the two sets of gluon 
field configurations from Table~\ref{tab:ensembles}. The top two rows give values for 
$u$ and $d$ masses when $m_u \ne m_d$; the third row gives the value 
when $m_u=m_d=m_l$ and is the same value as for $am_l^{\mathrm{sea}}$ in 
Table~\ref{tab:ensembles}. $am_s^{\mathrm{val}}$ takes the value 
given in Table~\ref{tab:ensembles}. 
The next three rows give $K$~masses in pure QCD, 
from fitting $\overline{s}l$, $\overline{s}d$ and $\overline{s}u$ correlation 
functions respectively. The final four rows give ratios showing the impact 
of quenched QED with $Q_u=2/3$ and $Q_d=1/3$.    
}
\begin{ruledtabular}
\begin{tabular}{lll}
 & VC-P & C-P \\ 
\hline
$am_u^{\mathrm{val}}$ & 0.00148 & 0.001157 \\
$am_d^{\mathrm{val}}$ & 0.00322 & 0.002523 \\
$am_l^{\mathrm{val}}$ & 0.00235 & 0.00184 \\
\hline
$am_K$ & 0.37902(22) & 0.30387(11) \\
$am_{K^0}^{u\neq d}$ & 0.38125(18) & 0.30564(10) \\
$am_{K^+}^{u\neq d}$ & 0.37665(23) & 0.30219(14) \\
\hline
$R^0_{\mathrm{QED}}(am_{K^0})$ & 1.000263(64) & 1.000269(13) \\
$R^0_{\mathrm{QED}}(am_{K^+})$ & 1.003978(68) & 1.003869(57) \\
$R^0_{\mathrm{QED}}(am_{K^0}^{u\neq d})$ & 1.0002677(89) & 1.0002748(66) \\
$R^0_{\mathrm{QED}}(am_{K^+}^{u\neq d})$ & 1.003976(75) & 1.003792(43) 
\end{tabular}
\end{ruledtabular}
\label{tab:Kqed}
\end{table}

\begin{table}
\caption{Results for QCD $\pi$ meson masses in lattice units and 
QCD+QED/QCD ratios $R^0$ as determined on the VC-P and C-P gluon 
field ensembles from Table~\ref{tab:ensembles} and with valence 
$u$ and $d$ quark masses as in Table~\ref{tab:Kqed}. 
The top four rows give $\pi$ masses in pure QCD, 
from fitting $\overline{l}l$, $\overline{u}u$, $\overline{d}d$ 
and $\overline{u}d$ quark-line connected correlation 
functions respectively. The final six rows give ratios showing the impact 
of quenched QED with $Q_u=2/3$ and $Q_d=1/3$.    
}
\begin{ruledtabular}
\begin{tabular}{lll}
 & VC-P & C-P \\ 
\hline
$am_{\pi}$ & 0.10174(14) & 0.081467(60) \\
$am_{\pi^0_{uu}}^{u\neq d}$ & 0.08114(13) & 0.064905(59) \\
$am_{\pi^0_{dd}}^{u\neq d}$ & 0.11870(12) & 0.095033(69) \\
$am_{\pi^+}^{u\neq d}$ & 0.10189(13) & 0.081461(62) \\
\hline
$R^0_{\mathrm{QED}}(am_{\pi^0_{uu}})$ & 1.00099(14) & 1.00211(99) \\
$R^0_{\mathrm{QED}}(am_{\pi^0_{dd}})$ & 1.000228(37) & 1.00105(29) \\
$R^0_{\mathrm{QED}}(am_{\pi^+})$ & 1.03500(11) & 1.032458(95) \\
$R^0_{\mathrm{QED}}(am_{\pi^0_{uu}}^{u\neq d})$ & 1.00090(24) & 1.00283(15) \\
$R^0_{\mathrm{QED}}(am_{\pi^0_{dd}}^{u\neq d})$ & 1.000248(24) & 1.00121(44) \\
$R^0_{\mathrm{QED}}(am_{\pi^+}^{u\neq d})$ & 1.03474(12) & 1.03222(10) 
\end{tabular}
\end{ruledtabular}
\label{tab:piqed}
\end{table}

\begin{table}
\caption{Results for QCD $\eta_s$ meson masses in lattice units and 
QCD+QED/QCD ratios $R^0$ as determined on the VC-P and C-P gluon 
field ensembles from Table~\ref{tab:ensembles} and with valence 
$s$ quark mass as in Table~\ref{tab:ensembles}. 
These results are obtained from fitting $\overline{s}s$ 
quark-line connected correlation 
functions. 
}
\begin{ruledtabular}
\begin{tabular}{lll}
 & VC-P & C-P \\ 
\hline
$am_{\eta_s}$ & 0.52617(16) & 0.422821(82) \\
$R^0_{\mathrm{QED}}(am_{\eta_s})$ & 1.0002725(91) & 1.0002897(23) \\
\end{tabular}
\end{ruledtabular}
\label{tab:etasqed}
\end{table}

Table~\ref{tab:Kqed} shows our results for $K$ mesons. 
In the pure QCD calculation with $m_u \ne m_d$ 
we see that the $K^0$ (with a $d$ quark) has a higher mass than the 
$K^+$. $am_K$, which is the $K$ mass with $m_u=m_d$ is very close to the 
root-mean-square average of these two, as expected. 
The impact of quenched QED (as seen from $R^0$) 
is much larger for the $K^+$ than 
for the $K^0$, as expected. It does not vary strongly 
with the light quark mass or with the lattice spacing. 
Note that the QED effects given in the Table do {\it not} 
include finite-volume corrections. 

Table~\ref{tab:piqed} shows the equivalent of Table~\ref{tab:Kqed} 
for $\pi$ mesons. The $\pi^0$ mesons here are made, in the $u \neq d$ 
case, either from 
a $\overline{u}u$ combination of propagators or a $\overline{d}d$. 
Up to small corrections in $m_d-m_u$ from quark-line disconnected diagrams (that are not included here) the mass of $\pi^0$ as 
measured in experiments is given by the average
\begin{equation}
m_{\pi^0} = \frac{1}{\sqrt{2}} \left( m_{\pi^0_{uu}}^2 + m_{\pi^0_{dd}}^2 \right)^{1/2} .
\end{equation}
In pure QCD both $m_{\pi^0}$ and $m_{\pi^+}$ are close to $m_{\pi}$. 
Table~\ref{tab:piqed} also includes $R^0$ ratios for each version of 
the $\pi$ meson. Again we see that QED effects are much larger, as expected, 
for $\pi^+$ than $\pi^0$. For $\pi^0$ the accuracy with which $R^0$ can be 
determined is not as good for the lighter $u$ case compared to the heavier 
$d$. However, in all $\pi^0$ cases, $R^0$ is very close to~1. 

Finally, Table~\ref{tab:etasqed} gives the corresponding information 
for the $\eta_s$ meson that is the ground-state in pseudoscalar 
$\overline{s}s$ quark-line connected correlation functions.  
This is an electrically neutral meson and has only very small QED 
effects that can be very accurately determined. 

From the results in Tables~\ref{tab:Kqed},~\ref{tab:piqed} 
and~\ref{tab:etasqed} we construct the ratio of meson masses below 
from our QCD + quenched QED results
\begin{equation}
\label{eq:etas-qed-ratio}
\left. R_{m^{\mathrm{QCD+QED}}} = \frac{m_{\eta_s}^2}{m_{K^0}^2 + m_{K^+}^2 - m_{\pi^+}^2} \right|_{\mathrm{QCD+QED}} .
\end{equation}
We use $m_{\pi^+}$ here rather than $m_{\pi^0}$ because no quark-line disconnected correlators are then needed. For $m_{K^+}$ 
and $m_{\pi^+}$ we need to apply QED finite-volume corrections to the results. We do this using the analytic formulae through 
$\mathcal{O}(1/L_s^2)$~\cite{Hayakawa:2008an, Davoudi:2014qua, Borsanyi:2014jba}:
\begin{equation}
\label{eq:fvolanalytic}
m(L_s) = m(\infty) -\frac{Q^2\alpha_{\mathrm{QED}}\kappa}{2L_s}\left(1+\frac{2}{mL_s}\right) 
\end{equation}
Here $Q$ is the meson electric charge, $m$ its mass and $\kappa = 2.8373$. 
Note that the $1/L^2_s$ piece of the 
finite-volume correction to $m^2$ is mass-independent and 
cancels between the $K^+$ and $\pi^+$ terms in the denominator of Eq.~\eqref{eq:etas-qed-ratio}. 

We also construct the ratio of masses in pure QCD that was studied 
in~\cite{Dowdall:2013rya}
\begin{equation}
\label{eq:etas-qcd-ratio}
\left. R_{m^{\mathrm{QCD}}} = \frac{m_{\eta_s}^2}{2 m_{K}^2 - m_{\pi}^2} \right|_{\mathrm{QCD}} .
\end{equation}

We then take the ratio of the two ratios in Eqs.~\eqref{eq:etas-qed-ratio} and~\eqref{eq:etas-qcd-ratio}
to determine the $\eta_s$ mass in a world with strong isospin-breaking 
and quenched QED effects in terms of the $\eta_s$ mass determined in pure QCD. The ratio of ratios can be determined accurately because of correlations between the masses in QCD+QED and QCD as shown by the accurate values of $R^0_{\mathrm{QED}}$.
We obtain:
\begin{align}
\label{eq:rmvals}
\frac{R_{m^{\mathrm{QCD+QED}}}}{R_{m^{\mathrm{QCD}}}} = 
\begin{cases}
0.9986(4) &\mathrm{VC\text{-}P} \\
0.9979(4) &\mathrm{C\text{-}P}  \, ,
\end{cases}
\end{align}
with the uncertainties dominated by that from the small differences between $m_u=m_d$ and $m_u\ne m_d$ mass combinations in pure QCD. The
ratio of ratios for the two different lattice spacing values agree with each other to better than 0.1\% and both can be encompassed by taking a result of 0.998(1). 

We can then multiply  the value 
given in Eq.~\eqref{eq:etas-mass2} by this result to obtain the ratio 
of Eq.~\eqref{eq:etas-qed-ratio} in the continuum limit at infinite volume.  
Note that we do this rather than just taking our values on ensembles VC-P 
and C-P for the ratio of Eq.~\eqref{eq:etas-qed-ratio}. The reason for this 
is that there are QCD finite-volume and discretisation effects in our 
results on these sets. These will almost entirely cancel in 
the results of Eq.~\eqref{eq:rmvals} but they will be present in the 
individual ratios. 

We obtain a value for the ratio in Eq.~\eqref{eq:etas-qed-ratio} of
\begin{equation}
\label{eq:qed-ratio-val}
 R_{m^{\mathrm{QCD+QED}}} = 1.0043(64)(10) .
\end{equation}
The second uncertainty comes from the uncertainty in the ratio of ratios. 
The final step is to insert the correct experimental masses 
into the denominator of $R_{m^{\mathrm{QCD+QED}}}$\footnote{Note that the physical values of the denominators of $R_{m^{\mathrm{QCD}}}$ and $R_{m^{\mathrm{QCD+QED}}}$ are not the same. They differ by $\Delta_E(m_{\pi^+}^2-m_{\pi^0}^2)$, which is 0.17\% of the QCD+QED denominator. This explains why the ratio of ratios in Eq.~\eqref{eq:rmvals} is smaller than 1 while the $\eta_s$ mass takes the same physical value in QCD and QCD+QED. } to obtain a value for the 
$\eta_s$~mass in QCD + quenched QED. This value is 
\begin{equation}
\label{eq:m-etas}
m_{\eta_s} = 688.4(2.2)(0.3) \ \mathrm{MeV}, 
\end{equation}
which is almost identical to the previous determination in 
\cite{Dowdall:2013rya}. This shows that the $\pi$ and $K$ mass combinations 
taken there to allow for $m_u\ne m_d$ and QED effects were appropriate to  
cover their impact on the physical value of $m_{\eta_s}$.  

We conclude that the physical $\eta_s$ mass is unchanged when QED is included, given the definition of it in QCD used in~\cite{Dowdall:2013rya}. We can then tune the $s$ quark mass so that the $\eta_s$ mass takes the same value in both theories. From the change of the lattice $\eta_s$ mass when QED is included at fixed $am_s$ (i.e. $R^0_{\mathrm{QED}}(am_{\eta_s})$, see Table~\ref{tab:etasqed}) we can work out the retuning of the $s$ quark mass needed, as discussed in Section~\ref{sec:smass}.

\end{appendix}

\bibliographystyle{apsrev4-2}
\bibliography{paper}{}

\end{document}